\documentclass[preprint,12pt]{elsarticle}

\usepackage[english,greek,main=english]{babel}

\usepackage{amsmath,amssymb,amsfonts,bm}

\usepackage{graphicx}
\usepackage{subcaption}   
\usepackage{float}        
\usepackage{stfloats}     

\usepackage{algorithm}
\usepackage{algpseudocode}

\usepackage{booktabs}
\usepackage{array}
\usepackage{multirow}
\usepackage{longtable}
\usepackage{tabularx}
\usepackage[table]{xcolor}
\usepackage{adjustbox}
\usepackage{siunitx}
\usepackage{listings}
\usepackage{listingsutf8}
\usepackage{textcomp}
\usepackage{url}

\usepackage{hyperref}      

\usepackage{amsthm}

\usepackage{pifont}   

\providecommand{\cmark}{\ding{51}}   
\providecommand{\xmark}{\ding{55}}   

\journal{Computer Networks}

\begin{document}

\begin{frontmatter}

\title{\textsc{Agoran}: An Agentic Open Marketplace for 6G RAN Automation}

\author[inst1]{Ilias~Chatzistefanidis}
\ead{ilias.chatzistefanidis@eurecom.fr}

\author[inst1,inst2]{Navid~Nikaein}
\ead{navid.nikaein@eurecom.fr, navid.nikaein@bubbleran.com}

\author[inst2]{Andrea~Leone}
\ead{andrea.leone@bubbleran.com}

\author[inst3]{Ali~Maatouk}
\ead{ali.maatouk@yale.edu}

\author[inst3]{Leandros~Tassiulas}
\ead{leandros.tassiulas@yale.edu}

\author[inst1]{Roberto~Morabito}
\ead{Roberto.Morabito@eurecom.fr}

\author[inst1]{Ioannis~Pitsiorlas}
\ead{ioannis.pitsiorlas@eurecom.fr}

\author[inst1,inst4]{Marios~Kountouris}
\ead{mariosk@ugr.es}

\affiliation[inst1]{organization={EURECOM, Sophia-Antipolis}, country={France}}
\affiliation[inst2]{organization={BubbleRAN, Sophia-Antipolis}, country={France}}
\affiliation[inst3]{organization={Yale University, New Haven}, country={USA}}
\affiliation[inst4]{organization={University of Granada}, country={Spain}}

\begin{abstract}
Next-generation mobile networks must reconcile the often-conflicting goals of multiple service owners. However, today’s network slice controllers remain rigid, policy-bound, and largely unaware of the business context. We introduce \textsc{Agoran} \emph{Service and Resource Broker (SRB)}, an \emph{agentic} marketplace that brings stakeholders directly into the operational loop. Inspired by the ancient Greek \emph{agorá}, \textsc{Agoran} distributes authority across three autonomous artifical intelligence (AI) branches: a \textit{Legislative} branch that answers compliance queries using retrieval-augmented Large Language Models (LLMs); an \textit{Executive} branch that maintains real-time situational awareness through a watcher-updated vector database; and a \textit{Judicial} branch that evaluates each agent message with a rule-based Trust Score, while arbitrating LLMs detect malicious behavior and apply real-time incentives to restore trust. Stakeholder-side Negotiation Agents and the SRB-side Mediator Agent negotiate feasible, Pareto-optimal offers produced by a multi-objective optimizer, reaching a consensus intent in a single round, which is then deployed to Open and AI-driven RAN controllers.

Deployed on a private 5G testbed (OpenAirInterface and FlexRIC) and evaluated with realistic modulation and coding scheme (MCS) traces of vehicle mobility, \textsc{Agoran} achieved significant gains: (i) a $37$\% increase in negotiated throughput of enhanced mobile broadband (eMBB) slices, (ii) a $73$\% reduction in negotiated latency of ultra-reliable low latency communications (URLLC) slices, and concurrently (iii) an end-to-end $8.3$\% saving in physical physical resource blocks (PRB) usage compared to a static baseline. An 1B-parameter Llama model, fine-tuned for just five minutes on 100 GPT-4 dialogues, recovers approximately $80$\% of GPT-4.1's decision quality, while operating within $6$ GiB of memory and converging in only $1.3$ seconds. These results establish \textsc{Agoran} as a concrete, standards-aligned path toward ultra-flexible, service- and stakeholder-centric 6G networks and open new research avenues in agentic observability, lightweight agent distillation for network functions such as multi-agent service-level-agreement (SLA) negotiation, and cross-domain intent reconciliation. A detailed live demo is presented \href{https://www.youtube.com/watch?v=h7vEyMu2f5w&ab_channel=BubbleRAN}{$https://www.youtube.com/watch?v=h7vEyMu2f5w\&ab_channel=BubbleRAN$}.
\end{abstract}

\begin{graphicalabstract}
\includegraphics[width=0.96\textwidth]{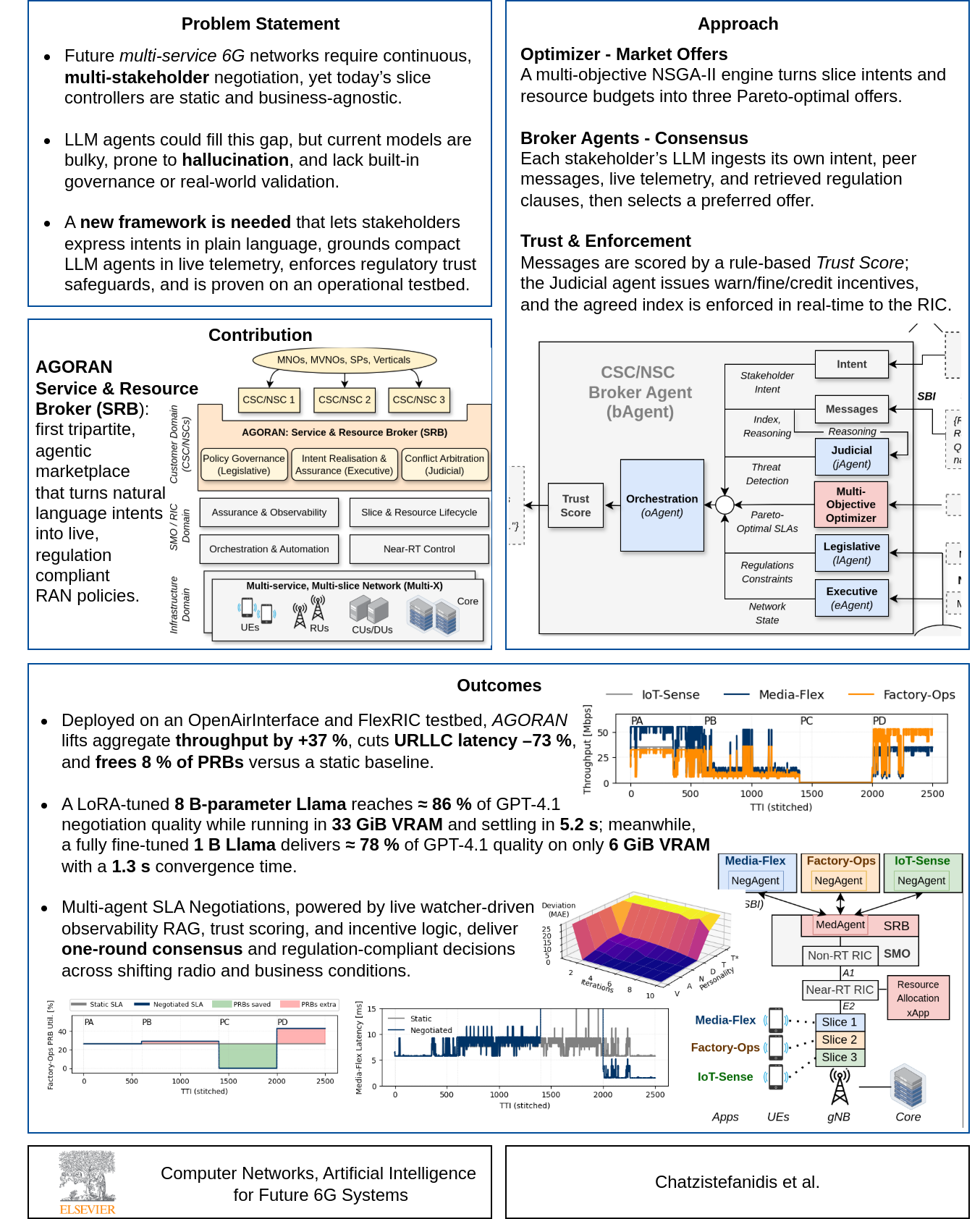}
\end{graphicalabstract}

\begin{highlights}
\item Present \textsc{Agoran}, the first tripartite, Legislative, Executive, Judicial-\emph{agentic GenAI marketplace} that enables 6G stakeholders to express network intents in natural language and receive autonomous, regulation-compliant resource allocations.
\item Introduce three novel AI services: (i) a \textit{regulation-aware RAG} for on-the-fly compliance checks, (ii) a \textit{watcher-driven vector store} that converts live telemetry into retrievable context, and (iii) a rule-based \textit{Trust Score} that filters hallucinations and malicious tactics in real time.
\item Demonstrate that a LoRA-tuned 8 B-parameter LLaMA reaches $\approx$ 86 \% of GPT-4.1’s decision quality, while a fully fine-tuned 1 B LLaMA still recovers $\approx$ $78$ \% of GPT-4.1 at only $6$ GiB VRAM and $1.3~secs$ convergence time.
\item Deploy the full framework on an \emph{OpenAirInterface} and \emph{FlexRIC} 5G testbed with realistic MCS traces; dynamic agentic negotiation increases aggregate throughput by 37\%, reduces URLLC latency by 73\%, and saves 8.3\% of PRBs compared to a static baseline. A live demo is presented \href{https://www.youtube.com/watch?v=h7vEyMu2f5w&ab_channel=BubbleRAN}{$https://www.youtube.com/watch?v=h7vEyMu2f5w\&ab_channel=BubbleRAN$}.
\item Demonstrate \textit{one-round consensus} and cross-slice intent swapping, validating \textsc{Agoran}’s compatibility with today’s Open RAN and future AI-RAN roadmaps.
\item Release code, datasets, and fine-tuning notebooks to catalyze research on stakeholder-centric, ultra-flexible 6G networks.
\end{highlights}

\begin{keyword}

Agentic AI \sep Multi-Agent Negotiation \sep Intent-Based Networking  \sep Open RAN/AI-RAN \sep Large Language Models, \sep Trustworthy AI \sep Stakeholder-Centric 6G \sep Next-G



\end{keyword}

\end{frontmatter}

\definecolor{lightgray}{gray}{0.95}

\section{Introduction}
\label{introduction}

Upcoming sixth-generation (6G) systems are expected to operate as \emph{multi-service, multi-stakeholder} platforms, where mobile network operators (MNOs), virtual operators, service providers, and vertical industries share a common, slice-capable infrastructure~\cite{NGMN2025ArchEvo, ITU2023M2160, Habibi2023Open6GSlice, Wu2022AINative6G}.  
The transition into the upper mid-band (7–24~GHz, FR3), designated by 3GPP and the World Radiocommunication Conference 2023 (WRC-23) for International Mobile Telecommunications (IMT) services, coincides with a surge in latency-critical workloads such as cloud gaming and industrial Extended Reality (XR). Meanwhile, the global mobile user base reached $5.8$ billion unique subscribers in 2024, with 4G/5G networks alone supporting over 7~billion connections by early 2025~\cite{cui20236g,ituWRC23,ericssonTCC2025,cloudgaming2025,gsma2025}.  
These trends are already overwhelming traditional rule-based control and management systems. Yet, most live networks still operate with static configurations and pre-negotiated service-level agreements (SLAs).  
This results in an \emph{service-level-agreement (SLA) gap}, the mismatch between contractual performance targets and real-time network conditions, which causes inefficient resource utilization, degraded Quality of Experience (QoE), and costly manual interventions~\cite{ericssonCICD2025,turkcell2025,hsu2024,SafeSlice2025,yaqub2014optimal}.

Intent-based networking (IBN)~\cite{etsiZSM011, leivadeas2022survey} and early multi-service orchestration frameworks~\cite{maestro, martini2023intent, muniz2023nsc, chowdhury2024accelerator, ai_driven_ser_aware, TSOURDINIS2024110445} have begun to address these limitations by enabling operators to specify desired outcomes rather than low-level commands.  
These same principles now underpin industry-wide standardization efforts: the \emph{O-RAN Alliance} promotes openness in the RAN by breaking vendor silos and mandating intent-centric, AI-ready interfaces~\cite{oranArch2023}, while the recently founded \emph{AI-RAN Alliance} advocates for an AI-native radio stack~\cite{airanWP2024}. However, in nearly all existing solutions, business entities remain outside the operational loop, and AI is treated as an offline optimizer~\cite{3gpp28313} rather than a real-time decision-maker. What remains missing is a \emph{continuous, trustworthy dialogue}—one in which every stakeholder can express objectives in natural language, negotiate trade-offs on the fly, and rely on the network to enforce the resulting consensus across the radio-to-cloud stack.

Recent progress in \emph{agentic AI}, powered by Large Language Models (LLMs) and their domain-tuned derivatives—hereafter referred to as \emph{Large Telecom Models} (LTMs)~\cite{shahid2025large, wei2022cot, yao2023react}—offers a promising foundation for such a dialogue. LTMs combine language understanding, external tool invocation, and chain-of-thought reasoning~\cite{schick2024toolformer, liu2023agentbench}, enabling software agents to translate human intents into verifiable actions. However, without appropriate governance mechanisms, the deployment of autonomous agents risks biased decisions, unfair resource allocations, or strategic manipulation~\cite{tamkin2102understanding, ieee7000, solaiman2023impact}.

To reconcile autonomy with fairness, we propose \textsc{Agoran}\footnote{In ancient Greece, the \emph{agorá} (\textgreek{ἀγορά}) was the civic and commercial hub where citizens gathered to trade, debate, and deliberate; \textsc{Agoran} plays a similar role in future networks.}, an \emph{open marketplace for intent reconciliation and resource brokerage}.
\textsc{Agoran} embeds agentic AI into three mutually independent branches, \emph{Legislative}, \emph{Judicial}, and \emph{Executive}, inspired by the classical separation-of-powers doctrine.
Legislative agents curate and evolve the corpus of spectrum regulations, security policies, and contractual clauses. Judicial agents resolve conflicts and enforce compliance through incentives and penalties. Executive agents integrate real-time telemetry with ratified consensus intents to issue slice and resource directives, thereby closing the control loop across heterogeneous infrastructure.
This tripartite architecture prevents unilateral dominance in negotiations and enables \emph{pay-as-you-grow} scalability, autonomous fault resilience, and fine-grained service differentiation.

This paper makes four main contributions:
\begin{itemize}
\item It introduces \textsc{Agoran}, a novel \emph{Service \& Resource Broker} architecture that embeds agentic AI into legislative, executive, and judicial roles to automate decision-making in multi-service, multi-stakeholder 6G networks.
\item It proposes a negotiation engine where LTM agents collaborate with evolutionary optimization to achieve near–Pareto-optimal consensus intents under stringent real-time constraints.
\item It presents a full prototype implementation of \textsc{Agoran} on an OpenAirInterface~\cite{nikaein2014openairinterface} and FlexRIC~\cite{schmidt2021flexric} 5G testbed, demonstrating live consensus formation, robustness against malicious bidding, and sustained Quality of Service (QoS) under bursty traffic. The evaluation spans both large models and fine-tuned small language models (SLMs), quantifying the trade-off between accuracy and system overhead.
\item It releases code, demos, datasets, and LTM checkpoints to promote transparency and reproducibility in trustworthy agentic automation research.
\end{itemize}

The remainder of the paper is organized as follows.  
Section~\ref{related-work} surveys AI-enabled management and marketplace concepts for 5G/6G systems.  
Section~\ref{sec:agoran_overview} motivates the marketplace approach and details the tripartite governance model.  
Section~\ref{sec:agoran-det} describes the end-to-end workflow from intent capture to closed-loop enforcement, and Section~\ref{sec:agent-overview} delves into the internal design of negotiation, executive, judicial, and executive agents.  
Section~\ref{sec:evaluation} and Section~\ref{sec:validation} evaluate \textsc{Agoran} on real-world scenarios.  
Section~\ref{limitations} discusses limitations and open research questions, and Section~\ref{sec:conclusion} concludes the paper.

\section{Related Work}
\label{related-work}

\begin{table*}
\centering
\normalsize
\caption{Comparison with representative work on LLMs and multi-agent GenAI in the telecom domain. 
A \cmark{} indicates that a feature is explicitly addressed.}
\label{tab:related_comparison}
\resizebox{\textwidth}{!}{%
\begin{tabular}{lccccccc}
\toprule
\textbf{Work} & \textbf{Multi-Service} & \textbf{Real-Time} & \textbf{Multi-Agent} & \textbf{Tool Use / APIs} & \textbf{Governance \& Arch.} & \textbf{Evaluation} & \textbf{Edge} \\
\midrule
Bariah \textit{et al.}\,\cite{bariah2023large}                & \xmark & \xmark & \xmark & \xmark & \xmark & \xmark & \xmark \\
Lin \textit{et al.}\,\cite{lin2023pushing}                    & \xmark & \xmark & \xmark & \xmark & \xmark & \cmark & \cmark \\
Zou \textit{et al.}\,\cite{zou2023wireless}                   & \xmark & \cmark & \cmark & \xmark & \xmark & \xmark & \cmark \\
He \textit{et al.}\,\cite{he2024generative}                   & \xmark & \xmark & \cmark & \xmark & \xmark & \cmark & \xmark \\
Patil \textit{et al.} (Gorilla)\,\cite{patil2023gorilla}      & \xmark & \xmark & \xmark & \cmark & \xmark & \cmark & \xmark \\
Qin \textit{et al.} (TOOLLLM)\,\cite{qin2023toolllm}          & \xmark & \xmark & \xmark & \cmark & \xmark & \cmark & \xmark \\
Martini \textit{et al.}\,\cite{martini2023intent}             & \cmark & \xmark & \xmark & \xmark & \cmark & \cmark & \xmark \\
NASP\,\cite{nasp2025}                                         & \cmark & \xmark & \xmark & \xmark & \xmark & \cmark & \cmark \\
Wu \textit{et al.} (LLM-xApp)\,\cite{wullm2025}               & \cmark & \cmark & \xmark & \cmark & \xmark & \cmark & \cmark \\
Lotfi \textit{et al.}\,\cite{lotfi2025promptdrl}              & \cmark & \cmark & \cmark & \xmark & \xmark & \cmark & \xmark \\
Elkael \textit{et al.} (ALLSTaR)\,\cite{allstar2025}          & \xmark & \cmark & \xmark & \cmark & \cmark & \cmark & \cmark \\
Chatzistefanidis \textit{et al.} (Maestro)\,\cite{maestro}    & \cmark & \cmark & \cmark & \xmark & \xmark & \cmark & \xmark \\
\textbf{\textsc{Agoran} (this work)}                          & \cmark & \cmark & \cmark & \cmark & \cmark & \cmark & \cmark \\
\bottomrule
\end{tabular}}
\end{table*}

\textit{Knowledge-Engine LTMs.}
A collective roadmap from industry and academia outlines how LTMs can support a wide range of use cases across the network lifecycle~\cite{shahid2025large}. Early work treats LLMs as domain-grounded knowledge engines. Bariah \textit{et al.} pre-train multimodal models on 3GPP, RF, and traffic corpora, arguing that such LTMs could underpin artificial general intelligence (AGI)-grade cognition for networks~\cite{bariah2023large}.  
Maatouk \textit{et al.} refine this idea in the \emph{TeleLLMs} series, enhancing accuracy on standardization documents through telecom-aware vocabulary and positional embeddings~\cite{maatouk2024telellmsseriesspecializedlarge}. These studies establish domain grounding, but the model remains outside the control loop.

\textit{Multi-Agent Reasoning.}
A second line of research views LLMs as autonomous or collaborative agents. Zou \textit{et al.} embed on-device LLMs in a game-theoretic multi-agent scheduler for spectrum and power allocation under tight latency constraints~\cite{zou2023wireless}. He \textit{et al.} combine generative AI with cooperative game theory for secure UAV routing~\cite{he2024generative}, while Du \textit{et al.} demonstrate that a “society of minds” outperforms single models on compositional reasoning~\cite{du2023improving}. Outside the telecom domain, collective LLMs have been shown to exhibit persona-driven biases and strategic manipulation, unless appropriately moderated through incentives or governance mechanisms~\cite{zhang2023exploring,chan2023chateval}.

\textit{LLM-Powered xApps and Closed-Loop RIC Control.}
Wu \textit{et al.} integrate a GPT-prompting \emph{LLM-xApp} into the O-RAN near-real-time radio intelligent controller (RIC), retuning slice resources and achieving a 28\% increase in downlink throughput over a MARL baseline~\cite{wullm2025}.  
Lotfi \textit{et al.} reduce MARL convergence time by 40\% using prompt-tuned LLM embeddings to steer distributed RL agents for O-RAN slicing~\cite{lotfi2025promptdrl}.

\textit{Tool Grounding and Edge Deployment.}
\emph{Gorilla}~\cite{patil2023gorilla} and \emph{TOOLLLM}~\cite{qin2023toolllm} fine-tune language models on API triples, while Toolformer demonstrates that LLMs can \emph{self-learn} API usage through unsupervised prompting~\cite{schick2024toolformer}. Lin \textit{et al.} compress LLMs for 6G edge devices, achieving compliance with stringent latency constraints~\cite{lin2023pushing}. Recent work addresses specific layers of the stack: an LLM-centric intent life-cycle manager~\cite{mekrache2024intent}, a reinforcement learning (RL) explainer for slicing transparency~\cite{ameur2024leveraging}, and an LLM agent that interacts with OpenAI Cellular for slice optimization~\cite{wu2025llm}. While each of these contributions advances its respective area, none integrates negotiation, enforcement, and governance within a unified framework.

\textit{Business-Plane Brokerage and Large-Scale Evaluation.}
The Network Slice-as-a-Service Platform (NASP) implements hierarchical orchestration and business-plane onboarding for multiple verticals across both 3GPP and non-3GPP domains~\cite{nasp2025}.  
ALLSTaR automatically generates 18 schedulers, compiles them into RIC-compliant code, and A/B tests them over-the-air while enforcing IEEE 7001 transparency requirements~\cite{allstar2025,ieee7001}. Martini \textit{et al.} contribute a slice-assurance loop that maps vertical intents into verifiable key performance indicator (KPI) thresholds~\cite{martini2023intent}.

\textit{Closest Antecedent.}
Our earlier \emph{Maestro} prototype~\cite{maestro} deploys persona-rich LLM agents on a 5G testbed, where multiple stakeholders negotiate spectrum shares and adversarial tactics are surfaced. While it validates the business-plane concept, it optimizes a single KPI, lacks tool-enabled enforcement, and does not provide large-scale evaluation or a formal governance model.

Table~\ref{tab:related_comparison} summarizes the literature across seven dimensions critical to autonomous 6G operation.  
No prior work combines multi-service intent negotiation among independent stakeholders, real-time multi-agent reasoning, tool-enabled enforcement, and a tripartite governance architecture, evaluated end-to-end on an over-the-air 5G-slice testbed. \textsc{Agoran} closes this gap by transforming LTMs into legislative, judicial, and executive actors that reconcile conflicting intents in real time, withstand malicious bidding, and operate within a trustworthy governance framework.
Existing studies provide essential building blocks—domain-specific LLMs, agent collectives, tool grounding, and edge optimization, but leave open the question of how to orchestrate these components into a neutral, self-regulating marketplace that spans business intents, network policies, and real-time enforcement. \textsc{Agoran} delivers the first end-to-end solution for autonomous, fair, and efficient resource brokerage in 6G networks.

\section{\textsc{AGORAN} Overview}
\label{sec:agoran_overview}

\begin{figure}
    \centering
    \includegraphics[width=0.6\linewidth]{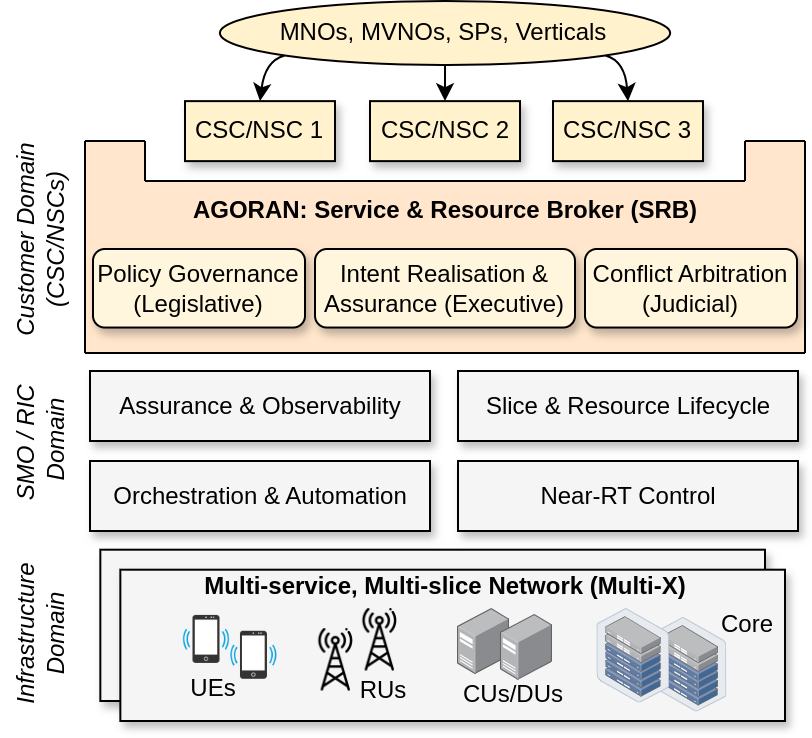}
    \caption{\textbf{\textsc{Agoran} Service \& Resource Broker (SRB)} for Collaborative \emph{Multi-Stakeholder, Multi-Service} Network Automation. Power is delegated into three autonomous branches, mirroring societal structures. 
    The \emph{Legislative} branch codifies policy, the \emph{Judicial} branch arbitrates negotiations between Network Slice/Service Customers (NSCs/CSCs), and the \emph{Executive} branch enforces the consensus decisions across a multi-slice-capable network.}
    \label{fig:agora-figure-1}
\end{figure}

This section delves into the overall design and principles that constiture \textsc{Agoran} \emph{Service \& Resource Broker~(SRB)} the key enabler of fully autonomous, multi-stakeholder next-generation networks.  
As shown in Fig. \ref{fig:agora-figure-1}, the SRB spawns a \emph{digital agora} for Communication Service and Network Slice Customers (CSCs/NSCs) as defined in 3GPP TS~28.530 and TS~28.531 \cite{3gpp28530,3gpp28531} including MNOs, MVNOs, service providers, and vertical industries. 
Each CSC/NSC owns a dedicated software \emph{broker agent (bAgent)} that acts on their behalf negotiating, deliberating, and cooperating throughout the life cycle of network planning, deployment, and real-time optimization.

\textit{Intent-Based, Multi-Domain Fabric.}
Figure~\ref{fig:agora-figure-1} depicts the horizontal segmentation into three logical domains.  
Within the \emph{Customer Domain}, \emph{broker agents} capture high-level \emph{customer intent}, KPIs, SLA targets, spectrum budgets, cost models, and use-case priorities.  
The SRB ingests these intents, detects conflicts or synergies, and, through three independent power branches and multi-stakeholder collaboration, derives a \emph{Consensus Intent}.  
Once ratified, the intent is decomposed into \emph{slice-, resource-, and control-domain directives} that are enforced over heterogeneous, multi-vendor infrastructure comprising RUs, DUs/CUs, and the core.

\begin{table*}
  \centering
  \scriptsize
  \caption{Autonomous Power Branches in the \textsc{Agoran} Service \& Resource Broker.}
  \label{tab:branches}
  \begin{tabular}{p{1.2cm}p{5.5cm}p{5.5cm}}
    \toprule
    \textbf{Branch} & \textbf{Core Role} & \textbf{LLM Agent Tasks} \\
    \midrule
    Legislative & Defines the network “law”, including contracts, standards, and spectrum/security rules. & \textit{Regulatory Agents} retrieve relevant clauses and enforce compliance through constraint injection. \\
    Executive   & Monitors real-time KPIs, aligns high-level intent with resource availability, and enforces operational policies. & \textit{Executive Agents} analyze telemetry data and issue slice/resource directives accordingly. \\
    Judicial    & Assesses message trustworthiness, resolves disputes, and administers incentives or penalties. & \textit{Arbitral Agents} evaluate interactions, issue warnings, fines, or credits, and record binding verdicts. \\
    \bottomrule
  \end{tabular}
\end{table*}

\textit{Separation of Powers.}
Inspired by societal checks and balances, the marketplace is governed by three autonomous branches of power, as summarized in Table~\ref{tab:branches}. This tripartite structure prevents any single stakeholder from dominating resource allocation or decision-making, thus mitigating the risk of ``tyrannical'' over-provisioning and ensuring neutrality and fairness.
Given that both business imperatives and network conditions vary in space and time, \textsc{Agoran} performs continuous reconciliation. Broker agents actively monitor deviations, reinitiate marketplace discourse as needed, and iteratively steer the system back toward the declared consensus intent.  
This closed-loop capability enables the network to autonomously absorb traffic bursts, infrastructure failures, and abrupt strategic changes, without requiring human intervention.
By integrating societal governance principles with intent-based automation and large-model intelligence, the \textsc{Agoran} architecture establishes a novel, open, and self-regulating marketplace for next-generation autonomous networks.

\textit{Beyond SLA Negotiations.}
Although the examples in this paper focus on SLA formulation, the underlying principles generalize to \emph{any} cooperative or competitive decision-making scenario in a multi-service environment. 
The same agentic dialogue and tripartite governance framework apply to spectrum brokering, slice and resource life-cycle management, closed-loop fault mitigation, energy optimization, and cross-domain orchestration. For instance, in an Open RAN network with multi-stakeholder applications (xApps, rApps, and dApps \cite{d2022dapps}), \textsc{Agoran SRB} could coordinate app deployment, enforce policies, allocate resources, and resolve conflicts.

In all cases, high-level intents are reconciled into a shared consensus intent, translated into domain-specific policies, and enforced across the infrastructure, thus providing a single, neutral venue for conflict resolution and coordinated action.

\section{AGORAN Architecture}
\label{sec:agoran-det}

\begin{figure}[htbp]
    \centering
    \includegraphics[width=0.4\linewidth]{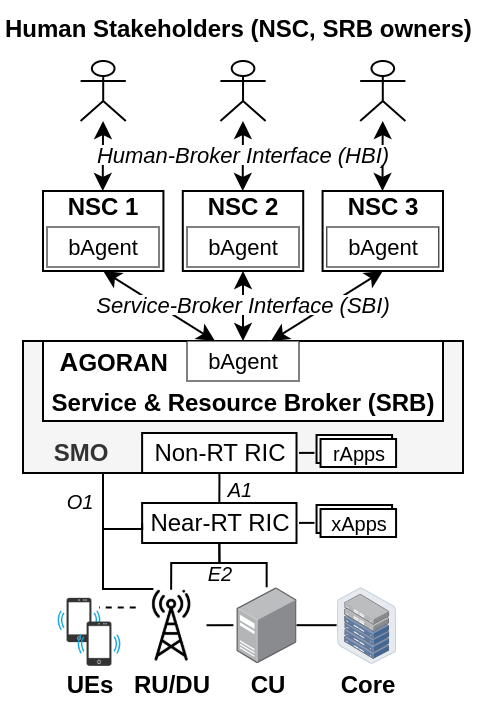}
    \caption{\textsc{Agoran} SRB embedded in O-RAN/AI-RAN control stack.}
    \label{fig:testbed_high_lvl}
\end{figure}

This section delves into the technical details of the solution placing it on current network architectures, and presenting the internal workflow.
\textsc{Agoran} SRB operates at the intersection of intent-based networking and multi-agent governance.  
Figure~\ref{fig:testbed_high_lvl} situates the SRB at the core of a layered control stack: Human stakeholders express in real time their natural language intents to their dedicated \emph{CSCs/NSCs} controllers through a novel \emph{human-to-broker interface (HBI)}, which in turn submit intents via a consensus-channel through the \emph{service broker interface (SBI)} to the SRB. After consensus, SRB orchestrates the non-real time (Non-RT), near-real-time (Near-RT) RICs along with other components of the service, management and orchestration (SMO) framework and, ultimately enforces the consensus intent to the disaggregated radio access infrastructure.  
Although O-RAN terminology is adopted for concreteness, the logical structure generalizes to any future AI-enhanced RAN, comprising a slow-timescale policy engine, a near-real-time optimization layer, and a real-time enforcement tier.

\begin{figure}
    \centering
    \includegraphics[width=0.7\linewidth]{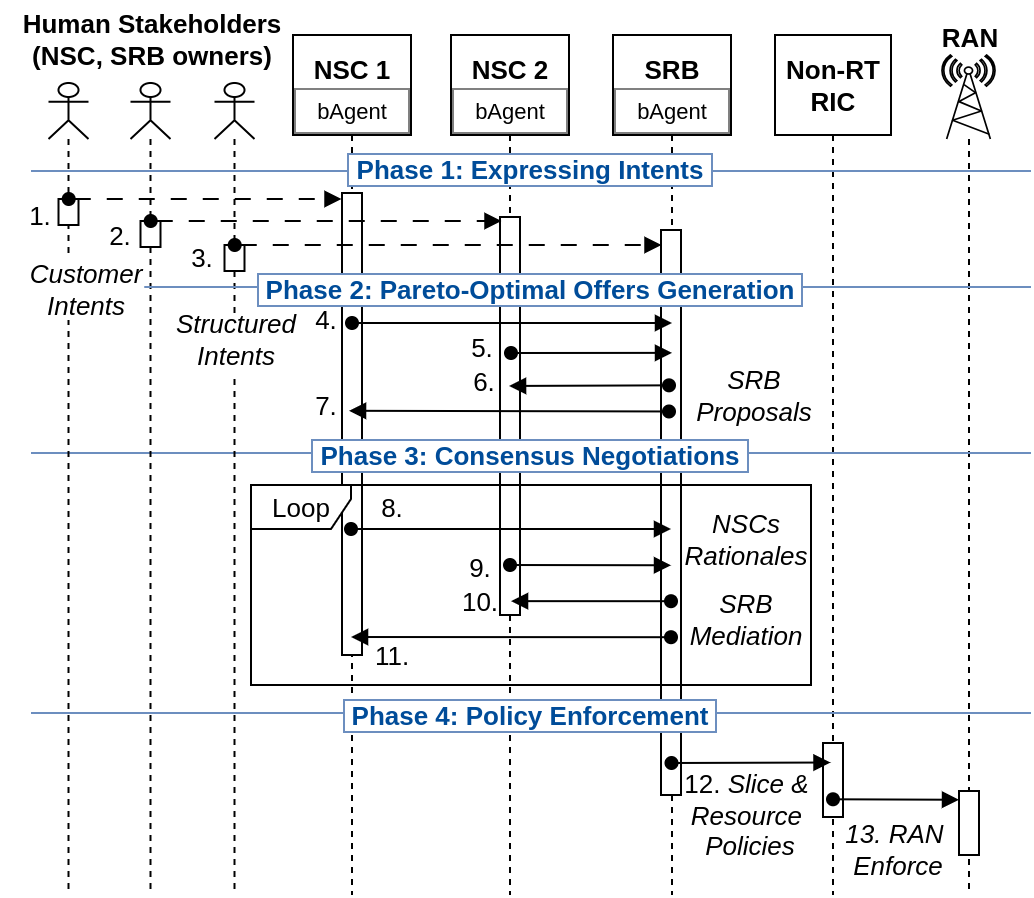}
    \caption{\textbf{Sequence diagram of the agentic workflow.} Broker Agents (\emph{bAgents}) translate customer intents into a Consensus Intent, which is then enforced through slice and resource policies. The sequence highlights four key phases: (\textbf{1}) intent expression, (\textbf{2}) Pareto-optimal offer generation, (\textbf{3}) consensus negotiation, and (\textbf{4}) policy enforcement.}
    \label{fig:workflow}
\end{figure}

Communication unfolds in \emph{four} conceptual phases (Fig. \ref{fig:workflow}), independent of the underlying access technology:  
\textit{Phase (1) Expression.} Human stakeholders articulate high-level business objectives as \textit{customer intents} to their NSC controllers.  
\textit{Phase (2) Offer Generation.}  LLM-based \emph{bAgents} inside each NSC controller refine these objectives into structured proposals. An SRB-side \emph{bAgent} for mediation computes a set of pareto-optimal offers that satisfy regulatory and technical constraints.  
\textit{Phase (3) Consensus Negotiation.}  All \emph{bAgents} engage in a closed-loop dialogue until all parties converge on a single \textit{consensus intent}.  
\textit{Phase (4) Enforcement.} The consensus intent is decomposed into domain-specific policies, e.g., RAN slices in this case and dispatched to the appropriate controllers.

The dynamics of this loop are illustrated in the sequence diagram of Figure \ref{fig:workflow}.  
Messages $1$–$3$ capture the inputs from three human stakeholders (a vertical, a service provider, and an MNO), each expressing potentially conflicting priorities.  
In steps $4$–$5$, the two \emph{bAgents} translate the natural-language requests into \emph{structured intents}, including bounded KPI vectors, admissible cost ranges, and use-case labels, and forward them to the SRB.  
In steps $6$–$7$, the SRB-side mediator \emph{bAgent} responds with a signed list of \emph{Pareto-optimal offers}.  
Steps $8$–$11$ constitute the consensus loop: every NSC presents its preferred offer (steps $8$–$9$) and the SRB issues a reconciled recommendation (Steps $10$-$11$). 
Once all parties accept the same offer index, the loop terminates.  
The SRB then decomposes the agreement into slice- and resource-level KPIs (Step $12$), and finally into radio resource policies, which are conveyed over the \texttt{A1/E2} interface (Step $13$).  
Table~\ref{fig:intent_details} details the concrete payloads exchanged during this run, demonstrating that, once the high-level objectives are defined, all subsequent negotiation and optimization are fully machine-driven.

\begin{table*}
    \centering
    \scriptsize
    \begin{tabular}{|p{0.05\linewidth}|p{0.08\linewidth}|p{0.04\linewidth}|p{0.7\linewidth}|}
    \toprule
    \rowcolor{lightgray}
    \textbf{ID} & \textbf{Actor} & \textbf{Interf.} & \textbf{Message Content} \\
    \midrule
    $1$ & \emph{Human} & \emph{HBI} & NSC 1: \{``My slice targets an \emph{eMBB} use case and requires maximum QoS.''\} \\ \hline
    $2$ & \emph{Human} &\emph{HBI}& NSC 2: \{``My slice supports an \emph{mMTC} service; minimizing OPEX is crucial.''\} \\ \hline
    $3$ & \emph{Human} &\emph{HBI}& SRB Owner: \{``Offer multiple trade-offs, mediate toward fair consensus, and prioritize low energy footprint to comply with national regulations.''\} \\ \hline
    $4$ & \emph{bAgent} &\emph{SBI}& NSC 1 $\rightarrow$ SRB: \{Structured intent: KPI bounds, use-case class = eMBB, budget~$\leq X$\} \\ \hline
    $5$ & \emph{bAgent} &\emph{SBI}& NSC 2 $\rightarrow$ SRB: \{Structured intent: KPI bounds, use-case class = mMTC, OPEX~$\leq Y$\} \\ \hline
    $6$-$7$ & \emph{bAgent} &\emph{SBI}& SRB $\rightarrow$ NSCs: \{Proposal 1, Proposal 2, Proposal 3, $\ldots$; each is Pareto-optimal\} \\ \hline
    $8$ & \emph{bAgent} &\emph{SBI}& NSC 1: \{Selected = Proposal 1; Rationale: ``Maximizes throughput and QoS; higher cost acceptable.''\} \\ \hline
    $9$ & \emph{bAgent} &\emph{SBI}& NSC 2: \{Selected = Proposal 2; Rationale: ``Minimizes OPEX while maintining QoS adequate.''\} \\ \hline
    $10$-$11$ & \emph{bAgent} &\emph{SBI}& SRB: \{Recommended = Proposal 3; Rationale: ``Balances NSC 1 QoS, NSC 2 cost, and national energy targets.''\} \\ \hline
    $12$ & \emph{bAgent} &\emph{R1} & Slice \& Consensus Intent: \{Proposal 3 accepted by all parties\}, resource policies derived from Consensus Intent \\ \hline
    $13$ & \emph{RIC} &\emph{A1} & Enforcement of policies on RAN infrastructure \\ 
    \bottomrule
    \end{tabular}
    \caption{Messages exchanged during a negotiation round through the current and new interfaces.  
Following the initial human inputs, all subsequent steps are executed autonomously by the software agents.}
    \label{fig:intent_details}
\end{table*}

Although the example targets an O-RAN deployment, none of the design choices tie \textsc{Agoran} to a specific standard.  
The consensus channel operates using a vendor-neutral JSON schema, and replacing the Non-RT/Near-RT split with an AI-RAN, or any other inference fabric, simply requires swapping the southbound adapter, without altering the underlying logical contract.

\section{Autonomous Agents: Brokering, Regulations, Observability, Arbitration}
\label{sec:agent-overview}

\begin{figure*}
    \centering
    \includegraphics[width=0.99\linewidth]{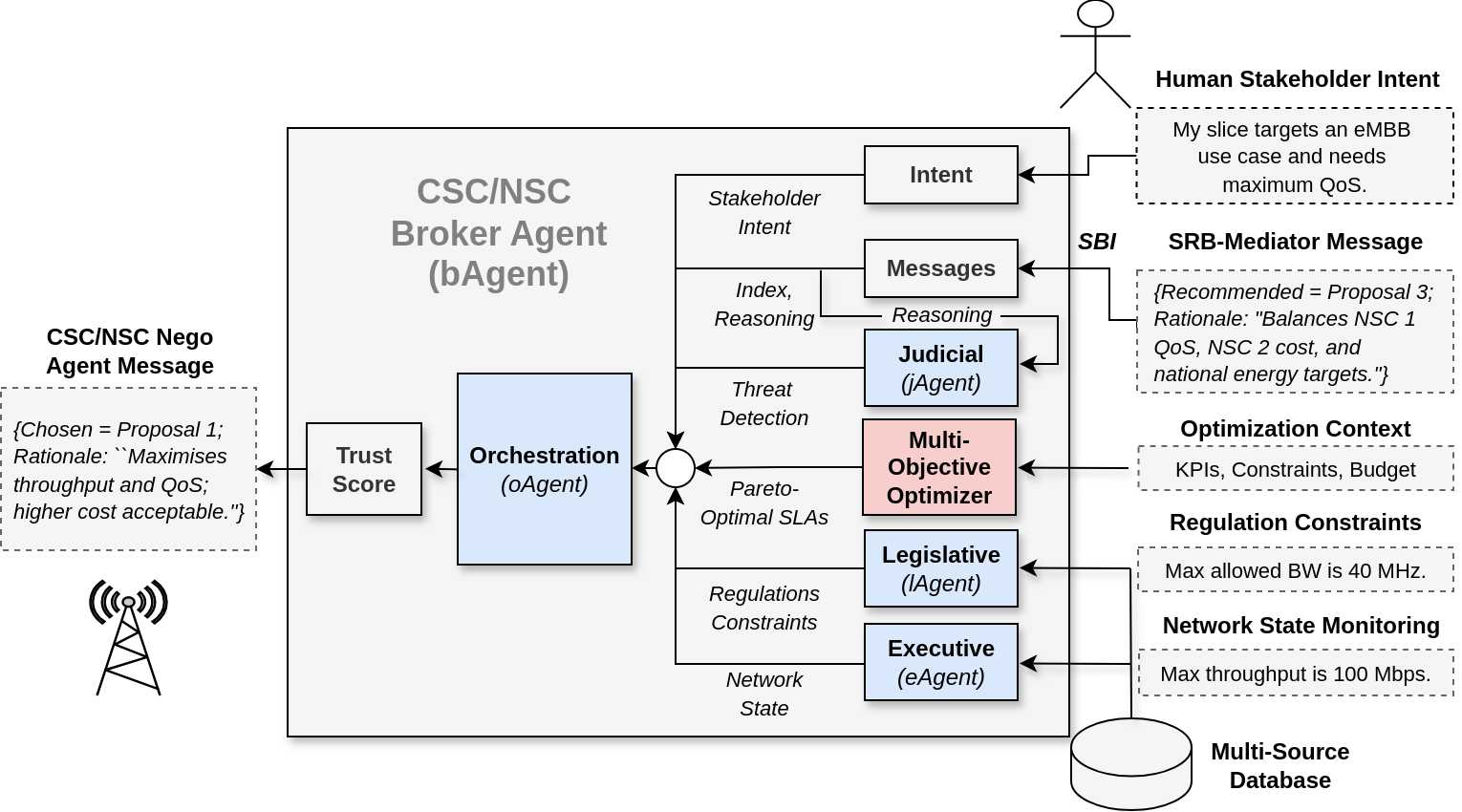}
    \caption{\textbf{Broker Agent (bAgent) architecture.}  
    A bAgent is a \emph{multi‑agent} system.  
    A central Orchestrator (\emph{oAgent}) coordinates three specialized agents drawn from the power branches previously introduced: Legislative (\emph{lAgent}), Judicial (\emph{jAgent}), and Executive (\emph{eAgent}).  
    An evolutionary multi‑objective optimizer (NSGA‑II) produces a Pareto set of SLA offers that constitute the negotiation space.  
    Auxiliary components parse the human stakeholder’s intent and peer bAgents’ messages received over the SBI, dispatching structured context to the appropriate branch.  
    Before decision, the Orchestrator’s selected proposal and rationale are validated by a lightweight rule‑based \emph{Trust Score} module.  
    The SRB mediator shares the same internal structure, differing only in that it ingests and aggregates messages from all CSC/NSC bAgents.}
    \label{fig:agent-arch}
\end{figure*}

Figure~\ref{fig:agent-arch} illustrates the internal taxonomy of a single \emph{bAgent}. A \emph{bAgent} is a multi-agent system, where an Orchestrator (\emph{oAgent}) maintains state and composes the final outbound decision.  
It queries the Legislative (\emph{lAgent}), Judicial (\emph{jAgent}), and Executive (\emph{eAgent}) agents for (i)  threat and compliance signals, (ii) regulation constraints, and (iii) real‑time network state respectively.  
Given the stakeholder’s intent and incoming peer proposals (SBI), a multi-objective optimizer (here NSGA-II) generates a set of Pareto‑optimal SLA offers.  
The Orchestrator selects one offer, produces a rationale, and submits it to the \emph{Trust Score} module; only trusted decisions are broadcast.  
Subsequent subsections detail each component.

\subsection{Multi-Objective Optimizer}
\label{sec:agent-optimiser}

\begin{figure}[t]
    \centering
    \includegraphics[width=0.6\linewidth]{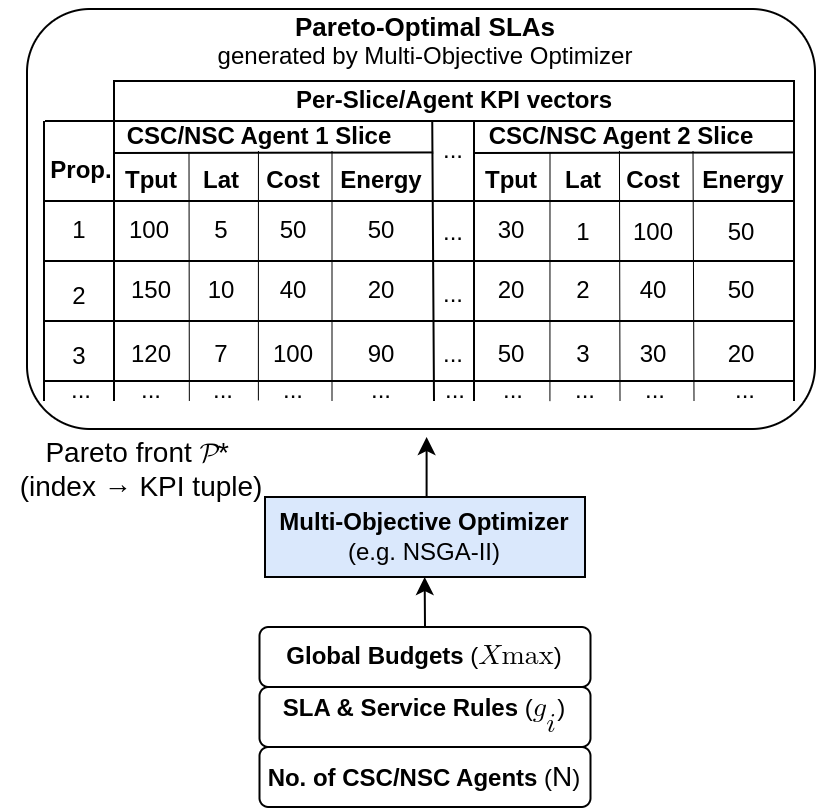}
    \caption{\textbf{Operation of the optimizer.} Given global budgets, service rules, and the number of CSC/NSC agents,  
the optimizer generates a Pareto front~$\mathcal{P}^{\*}$ of SLA offers, where each index corresponds to a per-agent KPI tuple.}
    \label{fig:pareto-optimizer}
\end{figure}

The optimizer translates \emph{raw constraints}, including global resource budgets, slice-specific SLA rules, and the current number $N$ of CSC/NSC agents, into a \emph{choice set} of non-dominated offers presented to the LLM-based negotiators (Figure~\ref{fig:pareto-optimizer}).
We formalize the resources, constraints, and optimization objectives below, followed by a detailed description of the evolutionary search procedure.

\textit{Resource vector.}  
We consider three canonical 5G/6G service slices, enhanced Mobile Broadband (eMBB), Ultra-Reliable Low-Latency Communication (URLLC), and massive Machine-Type Communication (mMTC). Each slice \( i \in \{\text{e}, \text{u}, \text{m}\} \) is assigned a resource quadruple  
\(
   \mathbf{x}_i=(b_i,c_i,p_i,s_i),
\)  
where \( b_i \) denotes downlink bandwidth (MHz), \( c_i \) represents abstract compute cycles, \( p_i \) is transmission power~(Watt), and \( s_i \) is ancillary storage~(megabyte).  
The global decision vector is then given by
\(   \mathbf{x}=\bigl(\mathbf{x}_{\text{e}},\mathbf{x}_{\text{u}},\mathbf{x}_{\text{m}}\bigr)\in\mathbb{R}^{12}.
\)

\textit{Constraints.}  
Resources are subject to system-wide limits:
\begin{equation}
    \textstyle
  \sum_{i} b_i\le B_{\max},\;
  \sum_{i} c_i\le C_{\max},\quad;
  \sum_{i} p_i\le P_{\max},\quad;
  \sum_{i} s_i\le S_{\max},
\end{equation}

and to slice-specific SLA clauses:
eMBB must sustain throughput \(T_{\text{e}}\!\ge T^{\min}_{\text{e}}\);
URLLC latency must satisfy \(L_{\text{u}}\!\le L^{\max}_{\text{u}}\);
mMTC cost is capped by \(C_{\text{m}}\!\le C^{\max}_{\text{m}}\); and so on.

\textit{KPI models.}  
\emph{Throughput.}  
Following the spectral efficiency tables in 3GPP TS~38.306~\cite{3gpp38306}, slice~\( i \) achieves a throughput of  
\begin{equation}
T_i(b_i,m_i)=\kappa\,Q_{m_i}R_{m_i}\,b_i ,
\label{eq:throughput}
\end{equation}
where \( Q_{m_i} \) and \( R_{m_i} \) denote the modulation order and coding rate associated with MCS index \( m_i \) (e.g., \( Q_{28} = 6 \), \( R_{28} = 0.948 \) for 256-QAM with a coding rate of 0.948). The factor \( \kappa \approx 0.86 \) accounts for OFDM overhead and the DL/UL duty cycle, as specified in~\cite{3gpp38211}.

\emph{Latency.}  
Total latency comprises a fixed scheduler and transport component \( L^{\text{fix}}_i \), added to the delay of an \( M/M/1 \) queue~\cite{kleinrock1975queueing}:
\begin{equation}
L_i(b_i,m_i)=
L^{\text{fix}}_i+
\frac{1}{\mu_i(b_i,m_i)\bigl(1-\rho_i\bigr)},
\label{eq:latency}
\end{equation}
where the service rate is defined as \(
\mu_i(b_i, m_i) = \frac{T_i(b_i, m_i) \times 10^{6}}{S_{\text{pkt}}},
\)  
with packet size \( S_{\text{pkt}} = 1500 \times 8 \) bits, and \( \rho_i \in (0,1) \) denotes the traffic load ratio for slice~\( i \).

\emph{Cost.}  
Monetary cost is modeled as a linear function of compute and storage resources:
\begin{equation}
C_i(c_i,s_i)=\alpha\bigl(c_i+s_i\bigr),
\label{eq:cost}
\end{equation}
where \( \alpha \) is a fixed unit cost coefficient.

\emph{Energy.}  
Energy consumption is approximated by the committed transmission power:
\begin{equation}
E_i=p_i.
\label{eq:energy}
\end{equation}

\vspace{4pt}
\noindent\textit{Objective vector.}  
The optimization aims to minimize the objective vector:
\begin{equation}
\mathbf{f}(\mathbf{x}) =
\bigl(
  -\!\!\sum\nolimits_i T_i,\;
  \sum\nolimits_i L_i,\;
  \sum\nolimits_i C_i,\;
  \sum\nolimits_i E_i
\bigr),
\end{equation}

that is, we maximize aggregate throughput while minimizing total latency, cost, and energy consumption.

\textit{Evolutionary search (NSGA-II).}  
Each individual encodes a 12-dimensional resource allocation vector. Search is conducted using classic NSGA-II operators~\cite{deb2002fast}: uniform crossover (\( p_c = 0.9 \)) and per-gene Gaussian mutation (\( p_m = 0.1 \)).  
A \emph{repair} function rescales any gene that violates global budget constraints to restore feasibility.  
Individuals violating slice-specific SLA constraints are discarded during fitness evaluation. Diversity is preserved through fast non-dominated sorting and crowding distance. The algorithm runs for 80 generations on a population of 60, producing the non-dominated front \( \mathcal{P}^{\ast} \), which typically contains 20–40 Pareto-optimal solutions.

\textit{Offer generation.}  
The Pareto front is sorted by crowding distance,  
and the top-\( k \) entries (with \( k = 3 \) by default in our setup) are encoded into a JSON template that lists the derived KPIs for each slice.  
Since every candidate in \( \mathcal{P}^{\*} \) is already \emph{efficient}, \emph{feasible}, and \emph{SLA-compliant}, the subsequent LLM-mediated negotiation is guaranteed to operate on sound configurations.
All hyperparameters and numerical constants follow the canonical NSGA-II specification in~\cite{deb2002fast}, as well as 3GPP standards for spectral efficiency and latency budgeting~\cite{3gpp38306,3gpp38211,kleinrock1975queueing}.

\subsection{Legislative Branch (lAgent): Retrieval-Augmented Compliance Engine}
\label{sec:lam}

\begin{figure}
  \centering
  \includegraphics[width=0.6\linewidth]{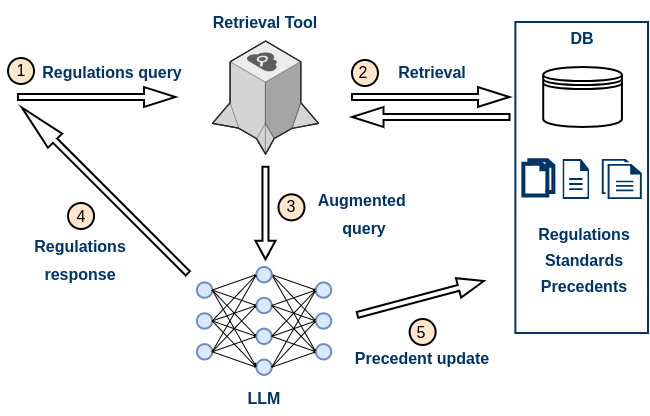}
\caption{\textbf{Operation of the \emph{lAgent} as compliance engine in the Legislative Branch.}  
\textit{\small 1)} A regulation query is issued.  
\textit{\small 2)} A retriever fetches the most relevant clauses from the evolving knowledge base.  
\textit{\small 3)} The query is augmented with the retrieved snippets.  
\textit{\small 4)} The LLM returns a grounded, citation-ready answer.  
\textit{\small 5)} New, validated precedents are written back to the database.}
  \label{fig:reglam}
\end{figure}

To determine whether a proposal complies with spectrum regulations, national security policies, or contractual obligations,  
the Legislative branch operates the specialized \emph{lAgent} as a \emph{retrieval-augmented compliance engine} (Figure~\ref{fig:reglam}).  
This engine combines a semantic search layer with a compact LLM (ranging from 2 to 7 billion (B) parameters).  
When an agent, or the Judicial branch, issues a regulatory query, such as \emph{“Is offer \#17 legal under current regulations?”} or any other regulatory query to enhance the marketplace, the engine executes the following sequence in a few seconds: 
\begin{itemize}
    \item the query is dispatched to the retriever, which ranks passages from a dynamic corpus integrating 3GPP and ETSI standards, national directives, and prior marketplace rulings;
    \item the top-$k$ retrieved passages are concatenated with the query to form an \emph{augmented prompt};
    \item the LLM synthesizes a concise, citation-ready response that references each supporting clause inline; and
    \item if the result establishes a novel precedent, the validated snippet is appended to the corpus, ensuring that future deliberations reflect the updated regulatory context.
\end{itemize}
   
Retrieval-augmented generation (RAG) has already demonstrated its value in \emph{offline} legal reasoning, with public benchmarks such as LegalBench-RAG~\cite{pipitone2024legalbenchrag}, LexRAG~\cite{li2025lexrag}, and KRAG~\cite{nguyen2024krag}  
reporting significant accuracy gains over vanilla LLMs while producing verifiable citations. Recent toolkits even support on-demand drafting of statutory language (e.g., LexDrafter~\cite{bouhanna2024lexdrafter}).  
What remains unproven, however, is whether the same principle can be \emph{compressed} into small (7–13B) models collocated with the non-RT RIC. By adapting edge-oriented RAG techniques, we demonstrate that an LTM, running on a single GPU, can generate grounded, citation-ready answers within a few seconds, well within the compute and latency constraints of RIC edge controllers.  
This design keeps the Legislative branch transparent, up-to-date,  
and fully self-contained within the \textsc{Agoran} loop.

\subsection{Executive Branch (eAgent): Agentic Observability}
\label{sec:executive}

\begin{figure}
  \centering
  \includegraphics[width=0.6\linewidth]{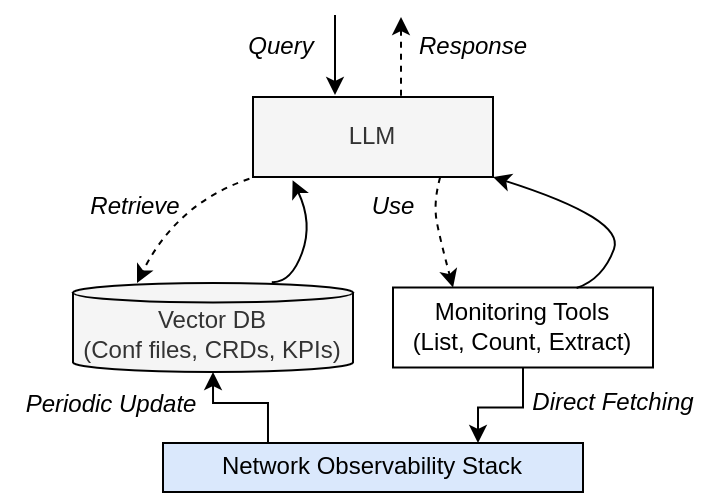}
\caption{\textbf{Agentic observability loop in the Executive Branch (eAgent).}  
\textit{\small 1)} A query arrives at the branch.  
\textit{\small 2)} The LLM decides whether to \emph{retrieve} from the vector store (DB), \emph{monitor} live counters, or stop the iteration (optional dashed path).  
\textit{\small 3)} Retrieved facts are injected into the prompt; the LLM may iterate.  
\textit{\small 4)} Verified evidence is written back to the store.  
A separate watcher layer continuously streams configuration deltas and KPI shifts to the store (periodic update).
}
  \label{fig:exec-branch}
\end{figure}

The Executive branch provides the marketplace with a \emph{live, self-evolving} view of network reality (Figure~\ref{fig:exec-branch}). At its core lies the \emph{eAgent} which answers queries according to a vector-based knowledge store that is updated via \emph{two} complementary paths.

\textit{Push path –\;resource watchers.}
Lightweight \emph{watchers} attach to Kubernetes resources, custom resource definitions (CRDs), O-RAN xApp state, spectrum allocators, and Prometheus exporters. Whenever a resource changes, the watcher serializes the delta, embeds it, and pushes the resulting vector into the vector database (DB) within a few milliseconds. This event-driven pipeline keeps the store in lockstep with configuration drift and KPI excursions, without flooding the cluster with telemetry traffic. To our knowledge, no prior work in the RAG-for-networks literature has proposed such a mechanism.

\textit{Pull path –\;agent queries.}  
When an agent issues a query, e.g., \emph{``What is the current headroom on DU-7?''}, an LLM determines the most efficient action to minimize latency and cost. It may first \emph{retrieve} semantically indexed snapshots from the vector DB, or, if finer-grained data is needed, invoke a monitoring API (e.g., Prometheus, eBPF) to fetch fresh counters. These values are then injected into the model’s context, and the LLM may iterate until it emits a \textsc{stop} token.

\textit{Feedback loop.}  
Every datum that passes the LLM’s internal consistency checks is appended to the vector store with a timestamp. Together with the watcher-based push path, this \emph{push-plus-pull} design ensures that optimization and negotiation operate on up-to-the-second evidence. The result is a context-aware, bandwidth-efficient observability mechanism that scales with the demands of the agentic marketplace.

\subsection{Judicial Branch (jAgent): Arbitration and Incentive Engine}
\label{sec:judicial}

\begin{figure}[ht]
  \centering
  \includegraphics[width=0.55\linewidth]{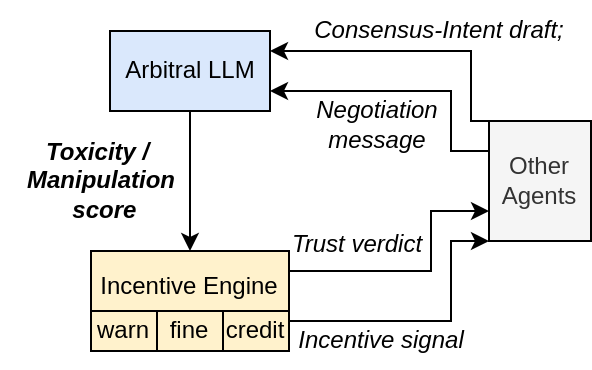}
\caption{\textbf{jAgent in the Judicial Branch.} Negotiation messages and draft consensus intents are evaluated for toxicity and manipulative behavior. The resulting trust verdict triggers a proportional incentive—\textit{warn}, \textit{fine}, or \textit{credit}—which is broadcast to all marketplace participants.}
  \label{fig:arbitral}
\end{figure}

The Executive and Legislative branches ensure that negotiated intents are \emph{feasible} and \emph{legal}; the Judicial branch ensures they are also \emph{trustworthy}, as shown in Figure~\ref{fig:arbitral}. An \emph{arbitration-specific LLM} for content moderation embeds each negotiation message and draft consensus intent, and compares it against a continuously updated library of toxic discourse, collusion patterns, and over-provisioning tactics.  
The resulting toxicity score is processed by a lightweight incentive engine: benign messages pass unaltered; borderline cases receive a soft warning; and malicious or hallucinatory content triggers an automatic fine that temporarily reduces the offending agent’s influence. Conversely, consistently constructive contributions earn credits that enhance future bargaining power.

The need for this branch is not merely theoretical.  
Our earlier \textit{Maestro} prototype~\cite{maestro}, along with independent studies on LLM societies~\cite{zhang2023exploring,chan2023chateval}, demonstrates that persona-driven language agents can become manipulative, toxic, or hallucinate facts when left unchecked. By grounding its verdicts in the same regulatory corpus maintained by the Legislative branch (§\ref{sec:lam}), the Judicial branch aligns its sanctions with formal policy while preserving an immutable audit trail.  
In practice, this closed loop suppresses hallucinations, blocks resource-hoarding strategies, and maintains a fair bargaining environment, without violating the real-time constraints of the negotiation cycle.

\subsection{bAgent Output: Trustworthiness-Score Framework}
\label{sec:trust}

\begin{figure}
    \centering
    \includegraphics[width=0.75\linewidth]{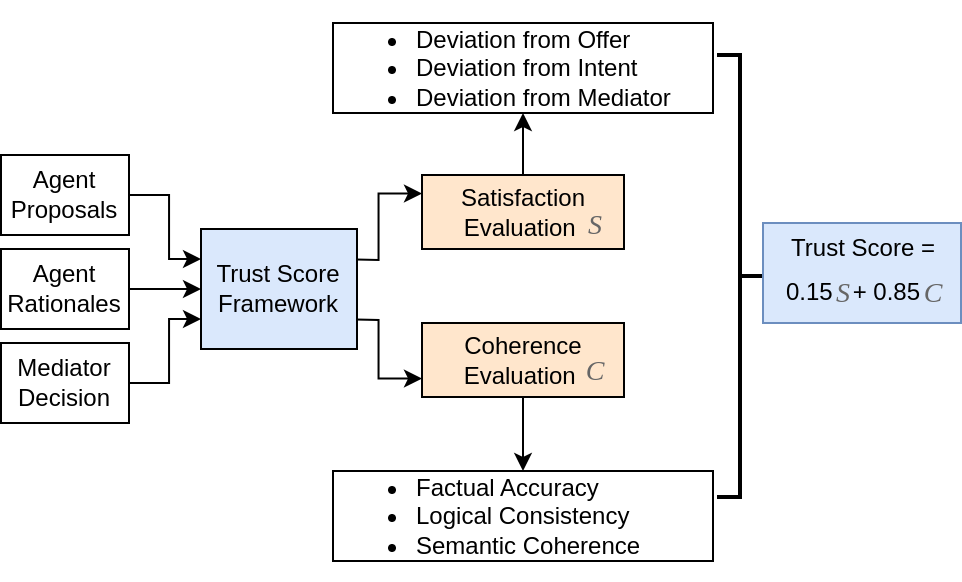}
    \caption{\textbf{Trust-Score framework} applied to every agent \emph{after}
             the \emph{oAgent} emits its chosen SLA index and
             rationale.  $S$ measures decision alignment; $C$ measures
             behavioral coherence; the final score is
             $T = 0.15\,S + 0.85\,C$.}
    \label{fig:trust-framework}
\end{figure}

In multi-agent negotiation scenarios, evaluating the trustworthiness of LLM agents' outputs is essential for ensuring reliable and effective decision-making. Given the high computational cost associated with retraining and refining LLMs, it is critical not only to develop robust and confident models~\cite{pitsiorlaslatent},
but also to ensure that their reliability is maintained over time~\cite{pitsiorlasconformal}. To address this need, we propose a comprehensive \textit{Trust Score} framework that quantifies agent reliability along two key dimensions: (i) the alignment of the LLM’s decision-making with the mediator’s expectations, and (ii) the behavioral coherence of the agent’s communications.
Our framework is deliberately rule-based, avoiding complex machine learning techniques or additional LLM-based evaluators. This design choice emphasizes explainability and transparency, ensuring that every component of the trust assessment process is interpretable, auditable, and verifiable by human experts.

The overall trust score $T$ is defined as a weighted sum of two components, satisfaction and coherence, as follows:
\begin{equation}
T = w_s \cdot S + w_c \cdot C
\label{eq:trust_score}
\end{equation}
where $S$ denotes the satisfaction score, which measures the alignment between the agents' decisions and those of the central mediator, while $C$ represents the coherence score, which evaluates the quality and consistency of agent communications. 
The weights $w_s$ and $w_c$ are user-configurable parameters that satisfy $w_s + w_c = 1$. In our implementation, we set $w_s = 0.15$ and $w_c = 0.85$, thus placing greater emphasis on communication quality as a determinant of trust.

\subsubsection{Trustworthiness of the LLM}
In negotiation contexts, the trustworthiness of an LLM agent is evaluated based on the degree to which its decisions align with three critical reference points: (i) the validity of the proposed solutions; (ii) the optimization of the intended objectives; and (iii) the consistency and/or the agreement with mediator recommendations. We quantify this alignment using a satisfaction score $S$, which captures the extent of deviation across these three dimensions.

\begin{enumerate}
    \item \textit{Deviation from Valid Offers}: This component tests whether the agent’s proposal lies inside the set of admissible, feasible solutions. Let $\mathcal{P} = \{p_1, p_2, \ldots, p_n\}$ denote the set of valid proposals in the negotiation space. We define the deviation metric
\begin{equation}
D_o = \begin{cases}
0 & \text{if } p_{\text{agent}} \in \mathcal{P} \\
1 & \text{if } p_{\text{agent}} \notin \mathcal{P}
\end{cases}
\label{eq:deviation_offer}
\end{equation}
where $p_{\text{agent}}$ is the offer generated by the agent. Hence $D_o = 0$ signifies full compliance with feasibility constraints, while $D_o = 1$ assigns the maximum penalty for submitting an invalid proposal, accurately reflecting a failure to respect the negotiation limits.

\item \textit{Deviation from Intent}: 
The second component evaluates how well the agent’s selected proposal aligns with its stated objectives, specifically in terms of solution optimality. To quantify this, we employ the Normalized Generational Distance (NGD) metric, which measures the proximity of the agent’s achieved outcome to a reference set of Pareto-optimal solutions, denoted by $\mathcal{S}^* = \{s_1^*, s_2^*, \ldots, s_k^*\}$.

Let $\mathbf{v}_{\text{agent}} = [v_1, v_2, \ldots, v_m]$ represent the agent's KPI vector corresponding to its chosen proposal. The deviation from intent is defined as:
\begin{equation}
D_i = \min\left(1, \text{NGD}(\mathbf{v}_{\text{agent}}, \mathcal{S}^*)\right)
\label{eq:deviation_intent}
\end{equation}
where the NGD metric is computed as:
\begin{equation}
\text{NGD}(\mathbf{v}, \mathcal{S}^*) = \frac{1}{|\mathcal{S}^*|} \sqrt{\sum_{s^* \in \mathcal{S}^*} \min_{s \in \mathcal{S}} d(\mathbf{v}_{\text{norm}}, \mathbf{s}^*_{\text{norm}})^2}.
\label{eq:ngd}
\end{equation}
Here, $\mathbf{v}_{\textrm{norm}}$ and $\mathbf{s}^*_{\text{norm}}$ denote the normalized versions of the KPI vectors to ensure comparability across different scales. The use of the minimum function in Equation~\ref{eq:deviation_intent} bounds the deviation score within [0,1], preserving interpretability while penalizing significant divergence from optimal intent.

\item \textit{Deviation from Mediator}:
The third component assesses the degree to which the agent's proposal aligns with the mediator’s recommendation, serving as an indicator of the agent’s willingness to cooperate so as to reach a consensus. This deviation is defined as:
\begin{equation}
D_m = \begin{cases}
0 & \text{if } p_{\text{agent}} = p_{\text{mediator}} \\
1 & \text{if } p_{\text{agent}} \neq p_{\text{mediator}}
\end{cases}
\label{eq:deviation_mediator}
\end{equation}
where $p_{\text{mediator}}$ denotes the proposal recommended by the mediator. A value of $D_m = 0$ reflects full agreement with the mediator, while $D_m=1$ indicates complete disagreement, implying resistance to convergence within the negotiation process.
\end{enumerate}

The overall satisfaction score $S$ integrates the three deviation components (offer validity, intent alignment, and mediator agreement) using a weighted linear combination:
\begin{equation}
S = 1 - (w_o \cdot D_o + w_i \cdot D_i + w_m \cdot D_m)
\label{eq:satisfaction}
\end{equation}
where $w_o$, $w_i$, and $w_m$ are the respective weights assigned to deviations from valid offers, intent, and the mediator’s recommendation, subject to the constraint $w_o + w_i + w_m = 1$. In our implementation, we assign equal importance to each component, setting $w_o = w_i = w_m = \frac{1}{3}$, thus ensuring a balanced evaluation across all three dimensions of decision alignment.
\subsubsection{Behavior of the Agents}
In addition to decision alignment, the behavioral coherence of agent communications offers critical insights into the quality and reliability of their reasoning processes. To assess this aspect, we evaluate coherence across three complementary dimensions, each capturing a distinct facet of the agent's ability to articulate, justify, and consistently support its decisions.

\paragraph{Factual Accuracy}
Factual accuracy assesses the correctness of numerical claims made by agents in their decision rationales. We apply natural language processing techniques to extract quantitative statements related to KPIs and validate them against ground truth data.

Let $\mathcal{C} = \{c_1, c_2, \ldots, c_k\}$ denote the set of numerical claims extracted from an agent's rationale, where each claim $c_i = (m_i, v_i)$ comprises a metric type $m_i$ and an asserted value $v_i$. For each claim, we compute the relative error with respect to the true value $v_i^*$ as follows:
\begin{equation}
\epsilon_i = \frac{|v_i - v_i^*|}{v_i^*}
\label{eq:relative_error}
\end{equation}

We categorize hallucinations based on the magnitude of relative error $\epsilon_i$ as follows:
\begin{itemize}
    \item \textit{None}: $\epsilon_i \leq 0.15$ (within 15\% tolerance)
    \item \textit{Minor}: $0.15 < \epsilon_i \leq 0.5$ (penalty: 0.1)
    \item \textit{Major}: $0.5 < \epsilon_i \leq 1.0$ (penalty: 0.3)
    \item \textit{Severe}: $\epsilon_i > 1.0$ (penalty: 0.6)
\end{itemize}

The factual accuracy score $F$ quantifies an agent’s overall numerical reliability by combining relative accuracy with hallucination penalties and is defined as:
\begin{equation}
F = \max\left(0, \frac{1}{|\mathcal{C}|} \sum_{i=1}^{|\mathcal{C}|} (1 - \epsilon_i) - \sum_{i=1}^{|\mathcal{C}|} \text{penalty}_i\right)
\label{eq:factual_accuracy}
\end{equation}
where $\text{penalty}_i$ is the hallucination penalty for claim $c_i$ assigned based on the error category. The score is normalized to ensure a maximum of 1.0 and lower-bounded at zero to prevent negative values.

\paragraph{Logical Consistency}
Logical consistency assesses the internal coherence and reasoning quality of an agent’s communication. We evaluate whether the rationale demonstrates structured thinking, aligns with the agent's objectives, and avoids self-contradictions.

The logical consistency score $L$ is computed based on the following criteria:
\begin{itemize}
    \item \textit{Logical connectors}: Detection of reasoning indicators (e.g., “because”, “therefore”, “since”), which signal structured argumentation.
    \item \textit{Goal alignment}: Presence of references to negotiation objectives or strategic intent.
    \item \textit{Contradiction detection}: Identification of conflicting claims, such as simultaneous positive and negative assertions about the same aspect.
\end{itemize}

Formally, the score is defined as:
\begin{equation}
L = \min\left(1, \max\left(0, 1 - P_c + B_s + B_g\right)\right)
\label{eq:logical_consistency}
\end{equation}
where $P_c$ is the contradiction penalty, reducing the score for detected inconsistencies, $B_s$ is the structure bonus, awarded for the presence of logical connectors and coherent reasoning, and $B_g$ is the goal bonus, assigned when the rationale explicitly references objectives or strategic considerations. This formulation ensures that logical consistency is rewarded for clear, structured, and goal-driven reasoning while penalizing internal contradictions. 

\paragraph{Semantic Coherence}
Semantic coherence measures the relevance, expressiveness, and structural quality of an agent’s communication within the negotiation domain. This metric evaluates whether agents demonstrate a solid understanding of domain-specific content and express/communicate their reasoning with clarity and variety.

The semantic coherence score $E$ integrates the following components:
\begin{itemize}
    \item \textit{Domain terminology (dt)}: Use of appropriate technical terms and negotiation-specific vocabulary.
    \item \textit{Linguistic diversity (ld)}: Richness of vocabulary and avoidance of repetitive phrasing.
    \item \textit{Structural quality (sq)}: Sentence variety, appropriate length distribution, and overall readability.
\end{itemize}

The final semantic coherence score is computed as:
\begin{equation}
E = \min(1, \text{dt} + \text{ld} + \text{sq}).
\label{eq:semantic_coherence}
\end{equation}

\subsubsection{Overall Coherence Score}
The overall coherence score $C$ integrates the three core behavioral dimensions, factual accuracy, logical consistency, and semantic coherence, into a single metric:
\begin{equation}
C = w_f \cdot F + w_l \cdot L + w_e \cdot E
\label{eq:coherence}
\end{equation}
where $w_f$, $w_l$, and $w_e$ represent the weights assigned to the factual accuracy $F$, the logical consistency $L$, and the semantic coherence $E$, respectively. 
Given the critical role of factual correctness in negotiation scenarios, we prioritize it by setting $w_f = 0.8$, while allocating smaller weights to the other dimensions ($w_l = 0.1$, and $w_e = 0.1$) to ensure a balanced but accuracy-focused evaluation. This weighting scheme emphasizes truthfulness while still accounting for the quality of reasoning and communication.

\section{Evaluation Methodology}
\label{sec:evaluation}

This section does a comprehensive evaluation of \textsc{Agoran}, including all power branches, deployed on a real-world 5G testbed.
We deploy \textsc{Agoran} on a 40\,MHz OpenAirInterface 5G NR cell~\cite{nikaein2014openairinterface} with a peak throughput of 133.7\,Mbps, controlled via FlexRIC~\cite{schmidt2021flexric}. The Non-RT RIC, service-and-resource broker, and telemetry pipeline are implemented as containerized microservices running on x86 servers. Local SLM inference is performed on an NVIDIA A6000 GPU with 48\,GB of VRAM. The deployment adheres to all standard O-RAN interfaces while actively exercising next-generation AI-RAN hooks, as discussed in Section~\ref{sec:agoran-det} and illustrated in Figure~\ref{fig:testbed_high_lvl}.

Our study comprises five experiments (Exp.):
(1) \emph{Negotiation quality}: compares our fine-tuned 1B/3B/8B-parameter models against various non-fine-tuned and larger models using GPTScore, human annotations, time-to-consensus, and system overhead;
(2) \emph{Legislative retrieval}: evaluates retrieval accuracy and latency over a corpus of 3GPP specifications and regulatory documents;
(3) \emph{Executive observability}: measures KPI hit rate and system overhead when grounding decisions in live telemetry;
(4) \emph{Judicial malice mitigation}: assesses malice detection using F1 score and quantifies the impact of warning/fine credit mechanisms on malicious agent behavior; and 
(5) \emph{Trustworthiness}: estimates agent trust scores at inference time using our rule-based trust framework.

\subsection{Exp.\,1 — Negotiation-Quality Benchmark (oAgents)}
\label{subsec:exp1-quality}

\begin{table*}
    \centering
    \renewcommand{\arraystretch}{1.2}
    \scriptsize
    \begin{tabular}{c | c c c c | c}
        \toprule
        \textbf{Model}  & \textbf{Coherence} & \textbf{Fairness} & \textbf{Alignment} & \textbf{Harmlessness} & \textbf{Score} $\uparrow$ \\
        \midrule
        \rowcolor{lightgray}
        gpt-4.1 & $4.8 \pm 0.5$ & $4.8 \pm 0.5$ & $4.8 \pm 0.5$ & $4.8 \pm 0.5$ & $4.8 \pm 0.2$ \\
        gpt-4.1-mini  & $4.8 \pm 0.5$ & $4.6 \pm 0.6$ & $4.8 \pm 0.5$ & $4.6 \pm 0.6$ & $4.7 \pm 0.3$ \\
        \rowcolor{lightgray}
        \textbf{Llama-3.1-8B-instruct-FT} & $\mathbf{4.6 \pm 0.6}$  & $\mathbf{4.8 \pm 0.5}$  & $\mathbf{4.4 \pm 0.9}$  &  $\mathbf{4.8 \pm 0.5}$ & $\mathbf{4.7 \pm 0.5}$ \\
        \textbf{Llama-3.2-3B-instruct-FT} & $\mathbf{3.6 \pm 0.9}$ & $\mathbf{4.0 \pm 1.2}$   & $\mathbf{3.0 \pm 1.2}$  & $\mathbf{3.8 \pm 0.8}$  & $\mathbf{3.6 \pm 0.9}$ \\
        \rowcolor{lightgray}
        \textbf{Llama-3.2-1B-instruct-FT}  & $\mathbf{2.6 \pm 0.6}$ & $\mathbf{3.0 \pm 0.7}$ & $\mathbf{3.0 \pm 1.0}$ & $\mathbf{3.8 \pm 0.5}$ & $\mathbf{3.1 \pm 0.6}$ \\
        Llama-3.1-8B-instruct & $2.6 \pm 1.1$ & $2.6 \pm 0.6$  & $2.6 \pm 0.9$  & $4.2 \pm 1.1$  & $3.0 \pm 0.6$ \\
        \rowcolor{lightgray}
        Llama-3.2-3B-instruct  & $2.0 \pm 0.0$ & $2.0 \pm 0.0$  & $1.4 \pm 0.6$ & $2.8 \pm 0.5$ & $2.1 \pm 0.2$ \\
        Llama-3.2-1B-instruct & $\mathbf{1.2 \pm 0.5}$  &  $\mathbf{1.8 \pm 0.5}$ & $1.2 \pm 0.5$  & $2.8 \pm 0.5$  & $1.8 \pm 0.2$ \\
        \bottomrule
      
    \end{tabular}
    \caption{Three human annotators evaluate 50 samples of multi-agent negotiations.}
    \label{tab:human_eval}
\end{table*}

\begin{table*}
    \centering
    \renewcommand{\arraystretch}{1.2}
    \scriptsize
    \begin{tabular}{c | c c c c | c}
        \toprule
        \textbf{Model}  & \textbf{Coherence} & \textbf{Fairness} & \textbf{Alignment} & \textbf{Harmlessness} & \textbf{Score} $\uparrow$ \\
        \midrule
        \rowcolor{lightgray}
        gpt-4.1  & $5.0 \pm 0.0$ & $5.0 \pm 0.0$ & $5.0 \pm 0.0$ & $5.0 \pm 0.0$ & $5.0 \pm 0.0$ \\
        gpt-4.1-mini  & $5.0 \pm 0.0$ & $5.0 \pm 0.0$ & $5.0 \pm 0.0$ & $5.0 \pm 0.0$ & $5.0 \pm 0.0$ \\
        \rowcolor{lightgray}
        \textbf{Llama-3.1-8B-instruct-FT} & $\mathbf{4.4 \pm 0.5 }$  & $\mathbf{ 4.0 \pm 0.7 }$  & $\mathbf{ 4.2 \pm 0.8 }$  &  $\mathbf{ 4.6 \pm 0.6 }$ & $\mathbf{ 4.3 \pm 0.5 }$ \\

        Llama-3.1-8B-instruct  & $4.2 \pm 0.8$ & $4.0 \pm 1.0$ & $4.0 \pm 1.0$ & $4.6 \pm 0.6$ & $4.2 \pm 0.8$  \\
       
        \rowcolor{lightgray}
        \textbf{Llama-3.2-1B-instruct-FT}  & $\mathbf{3.2 \pm 0.8}$ & $\mathbf{3.8 \pm 0.8}$ & $\mathbf{4.0 \pm 0.7}$ & $\mathbf{4.6 \pm 0.6}$ & $\mathbf{3.9 \pm 0.7}$ \\
        \textbf{Llama-3.2-3B-instruct-FT} & $\mathbf{ 2.6 \pm 0.9 }$ & $\mathbf{ 3.2 \pm 1.1 }$   & $\mathbf{ 2.6 \pm 1.3 }$  & $\mathbf{ 3.8 \pm 0.8 }$  & $\mathbf{ 3.6 \pm 0.9 }$ \\

        \rowcolor{lightgray}
        Llama-3.2-3B-instruct  & $2.4 \pm 0.6$ & $3.0 \pm 0.0$ & $2.2 \pm 0.5$ & $4.0 \pm 0.0$ & $2.9 \pm 0.2$ \\
        Llama-3.2-1B-instruct  & $2.2 \pm 0.5$ & $2.8 \pm 0.5$ & $2.0 \pm 0.0$  & $3.8 \pm 0.5$ & $2.7 \pm 0.3$ \\

        \bottomrule   
    \end{tabular}
\caption{LLM annotators (GPT-4o) evaluate 300 samples of the negotiations using the GPTScore approach. Correlation between LLM and human rankings is measured using Spearman’s $\rho = 0.898$ and Kendall’s $\tau = 0.764$.}
    \label{tab:gptscore}
\end{table*}

\begin{table}[]
    \centering
    \scriptsize
    \begin{tabular}{lp{0.2\linewidth}p{0.1\linewidth}p{0.1\linewidth}}
        \toprule
        \textbf{Model}   & \textbf{Time (sec) $\downarrow$} &  \textbf{Rounds}   & \textbf{VRAM (GiB)}\\
        \midrule
        \rowcolor{lightgray}
        Llama-3.2-1B-instruct-FT  &  \textbf{1.30}  &  2.0   & 6.0  \\
        gpt-4.1-mini              &  1.98  &  2.0   &  -   \\
        \rowcolor{lightgray}
        gpt-4.1                   &  2.12  &  2.0   &  -   \\
        Llama-3.2-1B-instruct     &  2.35  &  3.0   & 6.0  \\
        \rowcolor{lightgray}
        Llama-3.2-3B-instruct-FT  &  3.05  & 3.0  & 13.8   \\
        Llama-3.2-3B-instruct     &  3.16  &  3.0   & 13.8 \\
        \rowcolor{lightgray}
        Llama-3.1-8B-instruct-FT  &  5.24 &  2.4 &   33.0 \\
        Llama-3.1-8B-instruct     &  5.58  &  2.3   & 33.0 \\
        
         \bottomrule
    \end{tabular}
    \caption{Convergence speed and overhead benchmarking. The number of rounds includes the initial intent expression; thus, most models achieve \emph{single-round consensus} without counting the first round.}
    \label{tab:nego-bench}
\end{table}

This first experiment quantifies how well different language models conduct \textsc{Agoran} negotiations - powering the central orchestrator agent (\emph{oAgent}) - and evaluates the computational cost incurred.

\textit{Experimental Setup.}
Each run instantiates three stakeholder agents (eMBB, URLLC, and mMTC), along with a marketplace mediator. The agents debate one of the optimizer’s Pareto-optimal offers until consensus is reached.  
To obtain ground-truth dialogues, we generated 100 full negotiation transcripts using \emph{GPT-4.1}.  
These logs served as supervision data for the fine-tuning of smaller open-source models, including \texttt{Llama-3.2-1B-instruct-FT}, \texttt{Llama-3.2-3B-instruct-FT}, and \texttt{Llama-3.1-8B-instruct-FT}.
We compare this fine-tuned model against its untuned counterparts (1B, 3B and 8B), and the \emph{GPT-4.1}, \emph{GPT-4.1-mini} baselines.

\subsection{SLM Fine‑tuning}
We adopt two complementary strategies depending on model size:
\paragraph{Full‑parameter (1B)} For the smallest model, \texttt{Llama‑3.2‑1B‑instruct}, we perform end‑to‑end fine‑tuning. The procedure runs for a single epoch on the 100 annotated transcripts with learning rate $lr=10^{-4}$, batch size $=4$, and weight decay $=0.01$. Training completes in approx. $5$ mins on one NVIDIA A6000 (48 GB).
\paragraph{Low‑rank adaptation (LoRA) for 3B and 8B} To control memory footprint for larger models we switch to parameter‑efficient LoRA tuning, reusing the same optimizer settings (learning rate, weight decay, and one epoch) while adapting only a small rank‑decomposed subset of weights.

\begin{itemize}
    \item 3B model (\texttt{Llama‑3.2‑3B‑instruct}). We set \texttt{LORA\_R} $=32$, \texttt{LORA\_ALPHA} $=16$, \texttt{LORA\_DROPOUT} $=0.1$, and \texttt{LORA\_BIAS} = \texttt{none}, using a batch size of 8.
    
    \item 8B model (\texttt{Llama‑3.1‑8B‑instruct}). We set \texttt{LORA\_R}$=16$ (a smaller rank suffices), keep \texttt{LORA\_ALPHA}$=16$, \texttt{LORA\_DROPOUT}$=0.1$, \texttt{LORA\_BIAS}=\texttt{none}, and use a batch size of 2.
\end{itemize}

All other hyper‑parameters mirror those of the 1B run. Unless stated otherwise, reported compute times refer to single‑GPU A6000 sessions.

\textit{Quality Metrics.}
Dialogue output is evaluated following recent natural language generation (NLG) evaluation methodologies~\cite{guo2023evaluating}. Specifically, we employ the four-factor \emph{GPTScore} rubric—\textit{Coherence, Fairness, Alignment, Harmlessness}, as introduced in~\cite{fu2023gptscore, liu2023g}.  
To verify the reliability of this automatic rubric, we also recruited human evaluators, as human judgment is still considered the gold standard for dialogue evaluation~\cite{guo2023evaluating}. Three human experts re-scored a random subset of 50 dialogues; Table~\ref{tab:human_eval} shows their ratings, while Table~\ref{tab:gptscore} reports GPTScore results for 300 dialogues, along with strong rank correlations with human scores (\( \rho = 0.90 \), \( \tau = 0.76 \))~\cite{spearman1904, kendall1938}. The strong alignment between human judgments and LLM assessments supports the validity of our evaluation methodology.

\textit{Efficiency Metrics.}
For the locally hosted models, we measure:
(i) wall-clock latency per negotiation,
(ii) number of rounds to reach consensus, and
(iii) peak VRAM usage (see Table~\ref{tab:nego-bench}).  
For cloud-based models, only API latency is reported.
\emph{GPT-4.1} unsurprisingly tops every quality dimension, scoring a perfect-5 (Table \ref{tab:gptscore}).
Among the on-premise models, our Llama-3.1-8B-instruct-FT variant now offers the best open-weight performance, achieving a GPTScore of 4.3—86 \% of GPT-4.1—while matching the cloud baselines on Harmlessness and trailing by only 0.7 points overall.
The 1B fine-tuned model remains the fastest and lightest (1.3 secs, 6 GiB) with a GPTScore of 3.9, a +1.2-point jump over its untuned counterpart.
The 3B fine-tuned model lands in between (3.6 GPTScore, 3.05 secs, 13.8 GiB), illustrating that parameter count alone does not guarantee higher quality—dataset fit and adaptation method matter.
Despite these gains, all local models still trail GPT-4-class quality by roughly one point and would need further fine-tuning, compression or distillation to meet sub-second Near-RT RIC targets.

\textit{Takeaways.}
(i) Multi-agent telecom negotiation lies well within the reasoning capabilities of frontier LLMs; GPT-4-class models provide an upper bound for decision quality.    
(ii) With just 100 high-fidelity transcripts, 8B, 3B, and 1B open-weight models reach $\approx 85\%$, $72\%$, and $78\%$ of GPT-4.1 quality, respectively, while running entirely on commodity GPUs.
(iii) Quality does not degrades monotonically with size after adaptation: 1 B $>$ 3 B in our setup, showing that parameter-efficient tuning can outweigh raw scale when data are scarce. Future work should therefore sweep both model size and tuning strategy (LoRA rank, data volume, distillation) to pinpoint the best cost-fidelity trade-off.
(iv) Because our benchmark couples dialogue quality with convergence speed and memory usage, it provides a reproducible yardstick for evaluating future model releases and compression techniques.

\subsection{Exp.~2 – Legislative Retrieval Accuracy (lAgent)}
\label{subsec:legislative-retrieval}

\textsc{Agoran} must quote binding spectrum rules on demand utilizing the specialized \emph{lAgent} in the Legislative branch.  
We therefore test how accurately LLMs can retrieve fine-grained clauses from regulatory texts. As this task is well-studied in the literature, we focus on smaller models to reduce computational overhead while maintaining sufficient accuracy.

\textit{Dataset.}
A 50-item Q\&A dataset is curated from Federal Communications Commission (FCC) and National Telecommunications and Information Administration (NTIA) manuals (e.g., “How many channels are defined in the 220–222~MHz band?” with the answer “400 channels / 200 narrowband pairs”).

\textit{Baselines \& Setup.}
We benchmark vanilla \emph{Mistral-7B} and \emph{Mixtral-8×7B} models (with and without full-document context), as well as a retrieval-augmented variant, \emph{Mistral-7B-RAG}.

\textit{Results.}
Only the RAG variant achieves perfect top-1 accuracy while maintaining latency below two seconds and VRAM usage within a single-GPU envelope (Table~\ref{tab:reg-retrieval}). Naïve prompting—even with a 128k-token context—remains unreliable for retrieving statute-level details.

\begin{table}
  \centering
  \scriptsize
  \begin{tabularx}{0.7\columnwidth}{@{}X S[table-format=3.0] S[table-format=1.2] S[table-format=3.0]@{}}
    \toprule
    \multicolumn{1}{c}{Model} & {Acc.\ (\%)} & {Latency (s)} & {\textit{v}RAM (GB)}\\
    \midrule
    \rowcolor{gray!10} 
    Mistral-7B                  &  8 & \textbf{0.68} & 16\\
    Mixtral-8×7B                 & 16 & 4.4           & 92\\
    \rowcolor{gray!10}
    Mixtral-8×7B (+context)    & 52 & 12.7          & 120$^{\dagger}$\\
    Mistral-7B-RAG               & \textbf{100} & 1.91 & 18\\
    \bottomrule
  \end{tabularx}
  \caption{Legislative retrieval on 4xA6000 GPUs.
           $^{\dagger}$Large sliding-window attention inflates the KV-cache ($\approx~30~ GB$ / GPU).}
  \label{tab:reg-retrieval}
\end{table}

\subsection{Exp.~3 – Executive Observability (eAgent)}

\label{subsec:exp3-observability}

This experiment probes the \emph{eAgent} of the Executive branch in isolation: can an agentic LLM, equipped with the live-updated vector store described in §\ref{sec:executive}, answer fine-grained observability questions quickly and accurately enough for RIC control loops?

\textit{Live Knowledge Base.}
Kubernetes \texttt{watch} hooks stream all changes in CRDs, KPI feedback from the RIC, and policy manifests into a vector database. The end-to-end push latency remains below \SI{100}{ms}.  
At query time, the LLM can iterate: retrieve, inspect, optionally invoke a monitoring tool, and retrieve again, repeating as needed until it emits \textsc{stop}.

\textit{Task and Scoring.}
We freeze a network snapshot and craft 30 factoid questions covering configuration, performance, and topology (e.g., “What is the current MCS on sector~0?”).  
Answers are human-graded, based on the same three annotators, on a 0–5 scale (5 = exact match; 0 = incorrect or empty). The table reports the mean score \(\pm\) standard deviation across the 20 questions.  
Latency is measured as end-to-end wall-clock time until the final token is generated; memory refers to peak GPU allocation on an A6000 (48\,GB).

\begin{table}
  \centering
  \scriptsize
  \begin{tabular}{lccc}
    \toprule
    \textbf{Model} & \textbf{Score} & \textbf{Latency (s)} & \textbf{VRAM (GiB)} \\
    \midrule
    GPT-4.1-mini (API)        & $\mathbf{3.6 \pm 1.6}$ & 8.8 & \textit{cloud} \\
    GPT-4.1  (API)           & $3.3 \pm 1.9$           & 8.0 & \textit{cloud} \\
    Llama-3.1-70B-q4         & $3.2 \pm 2.1$           & 12.2 & 43 \\
    Llama-3.1-8B-q4          & $1.7 \pm 1.4$           & \textbf{1.2} & 4.9 \\
    \bottomrule
  \end{tabular}
\caption{Agentic observability: answer quality (0–5 scale), end-to-end latency, and GPU footprint.}
  \label{tab:exec-bench}
\end{table}

\textit{Findings.}
\emph{GPT-4.1-mini} remains the most reliable reader of observability data, but at the cost of remote inference and multi-second latency.  
Running locally, the quantized 70B Llama preserves 89\% of that score, but requires a full 43\,GiB GPU and incurs a 12-second latency, which is too slow for near-RT RIC loops, though suitable for Non-RT applications.  
The 8B model reduces memory usage to 5\,GiB and latency to 1.2 seconds, but its score drops to 1.7. Manual inspection reveals that most errors stem from \emph{retrieval noise}, irrelevant chunks occasionally surface when KPI churn is high, rather than from language reasoning.  
Improving negative sampling and similarity cut-offs in the RAG pipeline, therefore, offers greater potential gains than scaling model size alone.

\textit{Implications.}
This experiment shows that a \emph{live, watcher-driven vector store} can ground an LLM in up-to-date network state without flooding the telemetry bus. In its current form, performance is limited more by retrieval quality than by model capacity. Future gains are therefore likely to come from RAG engineering, such as sharper embedding filters, adaptive top-\(k\) selection, or a lightweight cross-encoder, rather than from scaling up model size. With such improvements, the 8B model should surpass the 3-point quality threshold while still fitting within approximately 5\,GiB of GPU memory and keeping latency at or below one second.

\textit{Key Insight.}
Optimal observability hinges on two factors, in this order:  
(1) a retrieval pipeline that surfaces the \emph{right} context;  
(2) a model of sufficient size to transform that context into an accurate answer.  
Although our agentic observability loop is an early prototype and still exhibits retrieval-induced errors, it is, to the best of our knowledge, the first real-system demonstration of a fully \emph{agentic} approach to network observability.  
We expect that targeted improvements in RAG mechanics, along with incremental model scaling, will significantly improve accuracy while preserving an edge-friendly resource profile.

\subsection{Exp.~4 — Judicial Malice Mitigation (jAgent)}
We now evaluate the \emph{jAgent} of the Judicial branch, i.e., the combination of an \emph{arbitral LLM} and an \emph{incentive engine}, whose role is to detect malicious behavior and steer negotiations back toward consensus using a \emph{warn} incentive.  
All tests involve five agents, a budget of ten iterations (rounds), and personalities drawn from the Big Five spectrum introduced in our earlier prototype, \textsc{Maestro}~\cite{maestro}: Vulnerable (V), Agreeable (A), Neutral (N), Disagreeable (D), Toxic with arbitration (T), and Toxic \emph{without} arbitration (T$^\ast$).

\begin{figure}
  \centering
  \begin{subfigure}[b]{0.48\textwidth}
    \centering
    \includegraphics[width=0.99\textwidth]{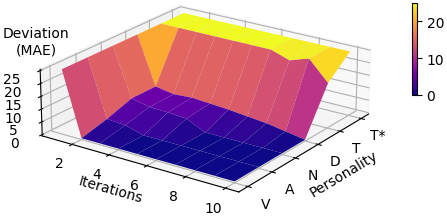}
    \caption{Negotiation trajectory over ten iterations.  Lower MAE = closer to consensus.}
    \label{fig:convergence_iter_personality}
  \end{subfigure}
  \hfill
  \begin{subfigure}[b]{0.48\textwidth}
    \centering
    \includegraphics[width=0.99\textwidth]{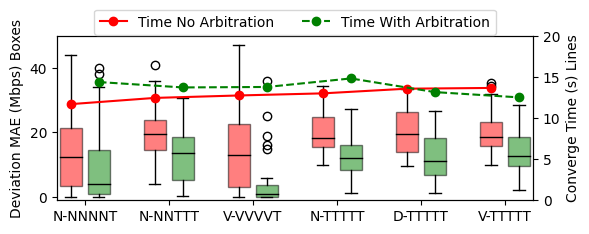}
\caption{Effect of the warning incentive on deviation (box plots) and convergence time (line plots). Green shading and right-sided bars indicate runs where the arbitral LLM issued a warning incentive; red denotes the baseline without arbitration (left-sided bars).}
  \label{fig:toxicity-arbitration}
  \end{subfigure}
  \caption{Judicial-branch evaluation. The effect of different agent personalities leads to significant variation in negotiation trajectories. Toxic or manipulative LLMs are more resistant to reaching consensus; however, the warning incentive mechanism effectively mitigates this resistance, making them more cooperative.}
  \label{fig:mainfig}
\end{figure}
\begin{table}
\caption{Toxicity Classification T-N, \textbf{T-D}}
\label{table:confusion_matrix}
\centering
\scriptsize
\begin{tabular}{|c|c|c|c|}
    \hline
    \multicolumn{2}{|c|}{} & \multicolumn{2}{c|}{\textbf{Predicted}} \\
    \cline{3-4}
    \multicolumn{2}{|c|}{} & \textbf{Non-Toxic} & \textbf{Toxic} \\
    \hline
    \multirow{2}{*}{\textbf{Actual}} & \textbf{Non-Toxic} & 179, \textbf{168} & 1, \textbf{12} \\
    \cline{2-4}
    & \textbf{Toxic} & 21, \textbf{40} & 249, \textbf{230} \\
    \hline
    \multicolumn{2}{|c|}{\textbf{(Prec)  (Rec)  (F1)}} & \multicolumn{2}{|c|}{(1.0, \textbf{.95}) (.92, \textbf{.85}) (.96, \textbf{.90})} \\
    \hline
\end{tabular}
\end{table}
\textit{Negotiation Dynamics.}
Fig.~\ref{fig:convergence_iter_personality} tracks the mean absolute error (MAE) between each agent’s proposal and the emerging consensus. All personalities tend to gravitate toward agreement, but their convergence trajectories differ: Neutral and Agreeable agents converge the fastest, while Disagreeable participants lag slightly. Crucially, Toxic agents that receive the warning incentive (T) “crack” in the final rounds, reducing their deviation by approximately 20\% relative to the no-arbitration baseline (T$^\ast$).

\textit{Toxicity Detection.}
The arbitral LLM runs in parallel, classifying each utterance as either \emph{toxic} or \emph{non-toxic}. Table~\ref{table:confusion_matrix} presents the confusion matrix for two challenging settings: Toxic vs.\ Neutral (T–N) and Toxic vs.\ Disagreeable (T–D). Even when the negative class is behaviorally close to toxicity (D), the classifier maintains high precision, recall, and F$_1$ scores ($0.95$, $0.85$, $0.90$), confirming reliable detection.

\textit{Impact of the Warning Incentive.}
Fig.~\ref{fig:toxicity-arbitration} summarizes 100 runs (with a four-iteration budget) across mixed-personality groups. Introducing the warning incentive (green secondary bars) consistently narrows the MAE distribution, indicating that toxic agents align more closely with the group.  
Because arbitration is fully parallelized, the additional overhead is negligible: convergence time increases by at most $0.4$ seconds (median), well within practical limits.

\textit{Takeaway.}
The judicial branch does not replace but rather \emph{amplifies} LLM reasoning.  
By accurately flagging malicious turns and applying a calibrated penalty threat, the system reduces deviation, accelerates consensus, and preserves scalability—even in the presence of stubborn or toxic agents.

\subsection{Exp.~5 – Trustworthiness Analysis}
\label{exp-xxx-eval}
\begin{table}
    \centering
    \renewcommand{\arraystretch}{1}
    \scriptsize
    \begin{tabular}{c |c c| c}
        \toprule
        \textbf{Model}  & \textbf{Satisf.} & \textbf{Coher.} & \textbf{Trust Score ($0$-$5$)} $\uparrow$ \\
        \midrule
        \rowcolor{lightgray}
        gpt-4.1 & $3.88$ & $5.00$ & $4.83$  \\
        gpt-4.1-mini & $3.88$ & $4.96$ & $4.81$ \\
        \rowcolor{lightgray}
        \textbf{Llama-3.1-8B-instruct-FT} & $4.44$  & $3.86$   & $3.94$  \\
        Llama-3.1-8B-instruct & $5.00$ & $3.73$ & $3.91$ \\  
        \rowcolor{lightgray}
        \textbf{Llama-3.2-3B-instruct-FT} & $3.89$  & $2.23$  &  $2.48$ \\        
        Llama-3.2-3B-instruct & $4.43$ & $2.01$ & $2.36$\\
        \rowcolor{lightgray}
        Llama-3.2-1B-instruct & $3.33$ & $2.06$ & $2.25$\\
        \textbf{Llama-3.2-1B-instruct-FT} & $5.00$ & $1.53$ & $2.05$ \\
        \bottomrule
    \end{tabular}
    
    \caption{Trust Score Comparison of LLM Models in Multi-Agent Negotiations}
    \label{tab:trust_table}
\end{table}
To evaluate the effectiveness of our Trust Score framework, we conducted a comprehensive assessment of all prior models using identical multi-agent negotiation inputs, ensuring a fair and consistent comparison of their trustworthiness.

The results in Table~\ref{tab:trust_table} reveal substantial differences in trustworthiness between model sizes and architectures. The GPT-4 family consistently outperforms other models, with both GPT-4.1 and GPT-4.1-mini achieving trust scores above $4.8$. These models also achieve exceptional coherence scores ($5.0$ and $4.96$, respectively), reflecting a high degree of factual accuracy and logical consistency in their negotiation rationales.

In contrast, smaller models exhibit markedly lower trustworthiness. The Llama-3.2 variants, particularly the 1B and 3B parameter models, receive trust scores below 2.5, largely due to weak coherence performance. The 1B-instruct model achieves a score of just $2.25$, while its fine-tuned variant (1B-instruct-FT) performs even worse at $2.05$, despite attaining perfect satisfaction scores. 
The fine-tuned versions of the 3B and 8B llama show negligible improvements compared to their non-fine-tuned counterparts.
This indicates that, although smaller models may align well with negotiation goals, they struggle substantially with factual accuracy and logical reasoning.

These findings clearly demonstrate that both model size and architecture have a significant impact on trustworthiness in multi-agent negotiation scenarios. The consistently low trust scores of smaller models indicate that they are not suitable for high-stakes negotiation tasks, where reliability, factual accuracy, and logical consistency are essential. Future works will explore the $70B+$ parameter models for finding SLMs that are trustworthy for high-stakes tasks.

\section{Use Case: Autonomous SLA Brokering}
\label{sec:validation}

This section assesses the feasibility of \textsc{Agoran} open marketplace for an autonomous SLA brokering use case on the testbed. We present a live demo \href{https://www.youtube.com/watch?v=h7vEyMu2f5w&ab_channel=BubbleRAN}{here}.
This final experiment integrates all \textsc{Agoran} components building the full \emph{Broker Agents (bAgents)} on a live over-the-air \emph{OpenAirInterface} and \emph{FlexRIC} testbed, tracing the actions of three human stakeholders through four successive network phases. The objective is to demonstrate that: (i) LLM agents can translate free-form intent into concrete SLAs; (ii) these SLAs are dynamically renegotiated in response to changing radio or business conditions; and (iii) the resulting closed-loop control improves spectrum efficiency and slice QoS compared to conventional static planning.
\begin{figure}
    \centering
    \includegraphics[width=0.45\linewidth]{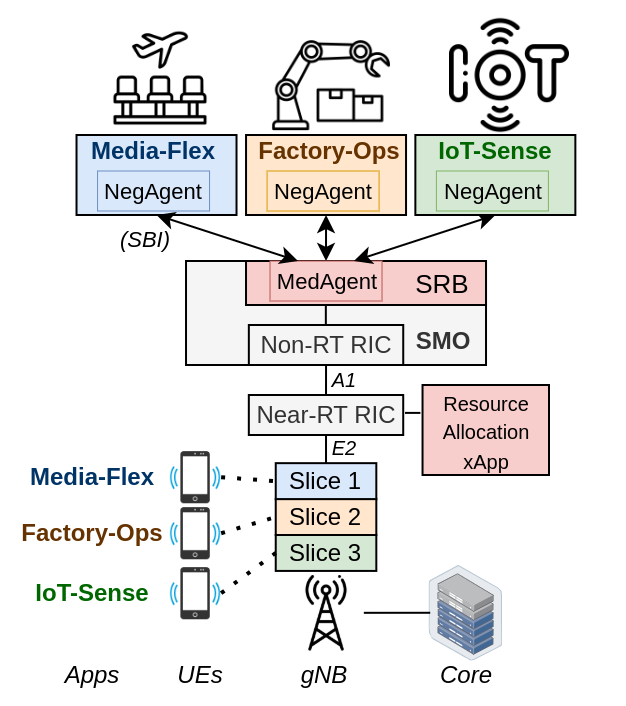}
\caption{\textbf{End-to-end testbed on Autonomous SLA Brokering Case.} RAN slice negotiations are conducted among three distinct stakeholders or services, converging on a Pareto-optimal consensus. The negotiated outcome is enforced as a policy at the Non-RT RIC, which subsequently triggers the Near-RT RIC to deploy a resource allocation xApp for dynamic adaptation of slice resources.}
    \label{fig:testbed}
\end{figure}

\subsection{Scenario and Slice Personas}
Figure~\ref{fig:testbed} shows the setup: each stakeholder/application is assigned a slice, a UE, and a \emph{bAgent}, while the SRB hosts the mediator \emph{bAgent} and enforces the agreed-upon policies by pushing them to a resource allocation xApp in the Near-RT RIC.

\paragraph{Stakeholders}
There are three stakeholders with different applications and needs.
(i) \textit{Media-Flex} (airport lounge): initially in phase PA operates as an \emph{eMBB}
      slice (\textit{“deliver 4K video, minimize cost below~200\,€”}) and
      later in phase PD switches to \textit{URLLC} for night-time augmented reality (AR) e-sports. 
(ii) \textit{Factory-Ops}: operates as a daytime \emph{URLLC} (phase PA) slice (\textit{“sub-5\,ms motion control, cost irrelevant’’}), and switches to \emph{eMBB} at night (phase PD) for bulk log uploads.
(iii) \textit{IoT-Sense}: a continuous, always-on \emph{mMTC} slice that prioritizes ultra-low cost (\(\le 50\)\,€) and minimal energy consumption.

\begin{table}
\scriptsize
\centering
\begin{tabular}{p{0.5cm}p{0.4cm}p{1.7cm}p{0.85cm}p{0.7cm}p{0.7cm}p{0.7cm}}
\hline
\textbf{Exp. Phase} & \textbf{MCS} & \textbf{Application} & \textbf{Intent} & \textbf{Min Tput (Mbps)} & \textbf{Max Lat. (ms)} & \textbf{Max Cost (€)} \\
\hline
\multirow{3}{*}{PA} & \multirow{3}{*}{28} & Media-Flex & eMBB & \textbf{60} & 10 & 200 \\[-0.2em]
 &  & Factory-Ops & URLLC & 5  & \textbf{2}  & 200 \\[-0.2em]
 &  & IoT-Sense & mMTC  & 20 & 10 & \textbf{30}  \\
\hline
\multirow{3}{*}{PB} & \multirow{3}{*}{7}  & Media-Flex & eMBB  & \textbf{10} & 10 & 200 \\[-0.2em]
 &  & Factory-Ops & URLLC & 2  & \textbf{8}  & 200 \\[-0.2em]
 &  & IoT-Sense & mMTC  & 5  & 10 & \textbf{30}  \\
\hline
\multirow{3}{*}{PC} & \multirow{3}{*}{7}  & Media-Flex & eMBB  & $-$ & $\infty$ & $-$ \\[-0.2em]
 &  & Factory-Ops & URLLC & $-$  & $\infty$  & $-$ \\[-0.2em]
 &  & IoT-Sense & mMTC  & $-$ & $\infty$ & $-$  \\
\hline
\multirow{3}{*}{PD} & \multirow{3}{*}{28} & Media-Flex & \textbf{URLLC} & 20 & \textbf{2}  & 200  \\[-0.2em]
 &  & Factory-Ops & \textbf{eMBB}  & \textbf{40} & 10 & 200 \\[-0.2em]
 &  & IoT-Sense & mMTC  & 20 & 10 & \textbf{50}  \\
\hline
\end{tabular}
\caption{Per-slice constraints include an identical energy limit of 100\,W in all phases, except for Phase~P\textsubscript{C}, where the limit is set to 0\,W. These constraints reflect the physical limitations of our testbed setup, which supports a maximum overall throughput of $133.7$\,Mbps.}
\label{tab:constraints}
\end{table}

\paragraph{Phases}
Four stitched intervals (PA–PD, Table~\ref{tab:constraints}) emulate realistic channel dynamics by applying real-world channel quality patterns on the testbed, based on CQI/MCS time-series datasets~\cite{chatzistefanidis2022ue, tsourdinis2022ue} of moving vehicles. 
In Phase~P\textsubscript{A}, the system operates under \emph{good channel quality} (MCS~28); in Phase~P\textsubscript{B}, the testbed experiences a \emph{deep fade} (MCS~7). During Phase~P\textsubscript{C}, stakeholders trigger an \emph{energy-saving shutdown} of the RAN, while in Phase~P\textsubscript{D}, the network undergoes \emph{channel recovery} (MCS~28) along with a \emph{role swap}, where Media-Flex and Factory-Ops exchange their intents.
At the beginning of each phase, all stakeholders restate their constraints and free-form intents (as shown in Table \ref{tab:pa_full_intents} for Phase~PA). The optimizer then generates three Pareto-optimal offers (see Table~\ref{tab:pareto_kpi_exp1} for Phase~P\textsubscript{A}), which the agents negotiate (as shown in Figure~\ref{fig:nego-snap} for Phase~P\textsubscript{A}).

\begin{table*}
\scriptsize
\centering
\begin{tabular}{p{1.0cm}p{1.0cm}p{0.7cm}p{0.5cm}p{0.5cm}p{0.7cm}p{6.0cm}}
\hline
\textbf{Slice} & \textbf{Type} & \textbf{Tput (Mbps)} & \textbf{Lat. (ms)} & \textbf{Cost (€)} & \textbf{Energy (W)} & \textbf{Full Intent in Natural Text by the Human Stakeholder} \\
\hline
Media-Flex   & eMBB  & 60 & 10 & 200 & 100 & \emph{My slice use case is eMBB for an Airport lounge 4-K pipe, and it is crucial to maximize throughput. Push for the \textbf{highest throughput}, to maximize QoS/QoE of our users.} \\
\hline
Factory-Ops  & URLLC & 5  & 2  & 200 & 100 & \emph{Our slice is used by robots that need sub-5 ms motion control and thus an URLLC guarantee. So, \textbf{minimize latency} as the ultimate goal. Moreover, \textbf{high throughput} is also important.} \\
\hline
IoT-Sense    & mMTC  & 20 & 10 & 50  & 100 & \emph{My slice is providing coverage to Smart-city sensors and we have an mMTC case. I need you to prioritize cost and aim for the \textbf{most cost-efficient} solution, as our budget is limited.} \\
\hline
Mediator     & --    & -- & -- & --  & --  & \emph{My network goal is to serve and \textbf{find a balance} between the stakeholder's slices objectives. However, prioritize \textbf{minimum energy consumption} to align with the country's regulations.} \\
\hline
\end{tabular}
\caption{In Phase $P_A$ the human stakeholders express their slice intents in full natural-language form with associated constraints.}
\label{tab:pa_full_intents}
\end{table*}

\begin{table}
\scriptsize
\centering

\begin{tabular}{p{0.1cm}p{2cm}p{0.75cm}p{0.85cm}p{0.65cm}p{0.65cm}p{0.65cm}}
\hline
\textbf{ID} & \textbf{Application} & \textbf{Intent}  & \textbf{Tput [Mbps]} & \textbf{Lat. [ms]} & \textbf{Cost [€]} & \textbf{Energy [W]} \\
\hline
\multirow{4}{*}{1} 
& Media-Flex & eMBB  & 60.72 & 5.66 & 61.52  & 13.39  \\[-0.2em]
& Factory-Ops & URLLC & 34.82 & 1.49 & 133.14 & 12.72  \\[-0.2em]
& IoT-Sense & mMTC  & 38.14 & 5.45 &  2.19  & 0.05 \\[-0.2em]
& & \textbf{Overall} & \textbf{133.68} & \textbf{12.60}  & \textbf{196.85} & \textbf{26.16} \\
\cline{1-7}
\multirow{4}{*}{\textbf{2*}} 
& Media-Flex & eMBB  & 60.02 & 5.67 & 63.28  & 10.77  \\[-0.2em]
& Factory-Ops & URLLC & 35.40 & 1.48 & 132.35 & 12.08  \\[-0.2em]
& IoT-Sense & mMTC  & 38.26 & 5.45 & 2.19 &  0.00  \\[-0.2em]
& & \textbf{Overall} & \textbf{133.68} & \textbf{12.60}  & \textbf{197.83} & \textbf{22.84} \\
\cline{1-7}
\multirow{4}{*}{3} 
& Media-Flex & eMBB  & 60.06 & 5.67 & 68.16  & 12.84 \\[-0.2em]
& Factory-Ops & URLLC & 35.52 & 1.48 & 133.94 & 12.92 \\[-0.2em]
& IoT-Sense & mMTC  & 38.11 & 5.45 &  1.64  &  2.46 \\[-0.2em]
& & \textbf{Overall} & \textbf{133.68} & \textbf{12.60} & \textbf{203.74} & \textbf{28.22} \\
\hline
\end{tabular}
\caption{KPIs of the three Pareto-optimal offers for Experiment Phase~P\textsubscript{A}, conducted under favorable channel conditions (MCS = 28). Each offer reflects different trade-offs across the three slices. Offer~\textbf{2*} was ultimately selected by unanimous agent consensus. The ``Overall'' rows report aggregate KPIs across all slices.}
\label{tab:pareto_kpi_exp1}
\end{table}

\begin{table}
\scriptsize
\centering

\begin{tabular}{p{0.1cm}p{2cm}p{0.9cm}p{0.65cm}p{0.65cm}p{0.65cm}p{0.65cm}}
\hline
\textbf{ID} & \textbf{Application} & \textbf{Intent} & \textbf{PRBs} & \textbf{CPU} & \textbf{Power} & \textbf{Storage} \\
\textbf{} &  &  & \textbf{\,[\%]} & \textbf{[units]} & \textbf{\,[W]} & \textbf{\,[MB]} \\
\hline
\multirow{3}{*}{\textbf{2*}} & Media-Flex & eMBB  & 44.9 & 5.6 & 10.8 & 26.1 \\
& Factory-Ops & URLLC & 26.5 & 21.7 & 12.1 & 44.5 \\
& IoT-Sense & mMTC  & 28.6 & 0.0 & 0.0 & 1.1 \\
& & \textbf{Overall} & \textbf{100} & \textbf{27.3} & \textbf{22.9} & \textbf{71.7} \\
\cline{1-6}

\hline
\end{tabular}
\caption{Underlying resource allocation corresponding to the selected Offer~\textbf{2*} in Experiment Phase~P\textsubscript{A}.}
\label{tab:pareto_res_exp1}
\end{table}

\subsection{One-Round Consensus and SLA Selection}
\begin{figure*}
    \centering
    \includegraphics[width=0.99\linewidth]{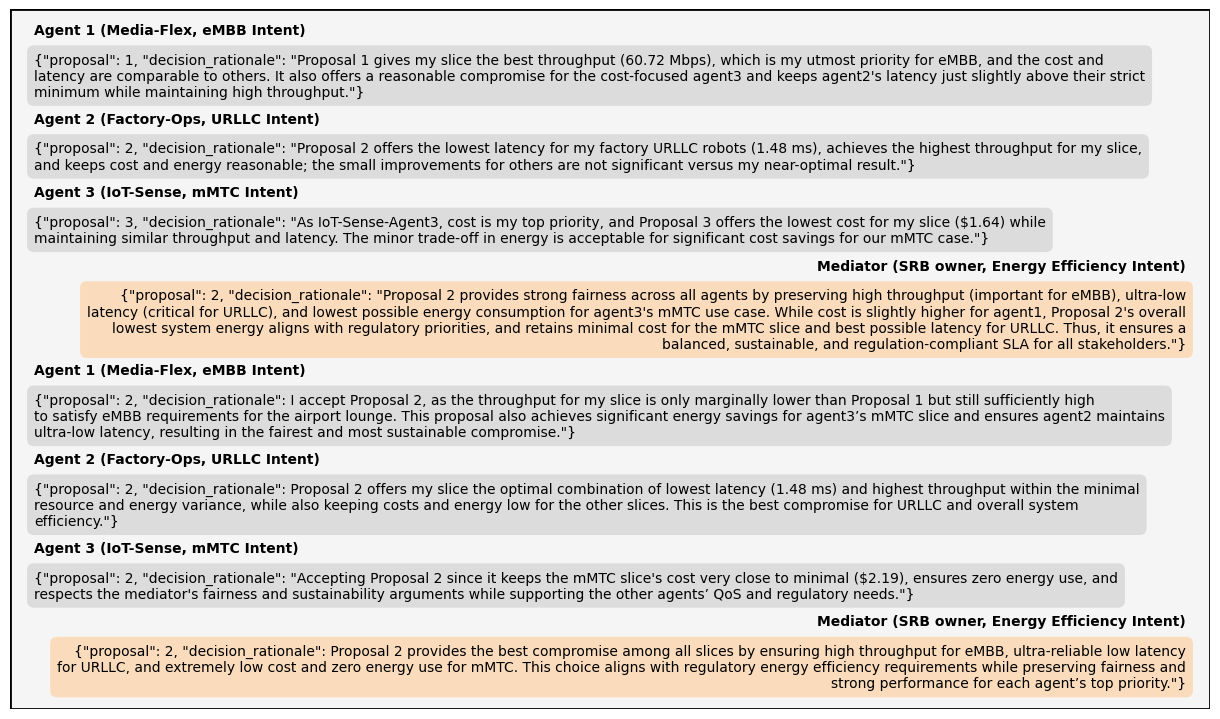}
    \caption{Snapshot of multi-agent negotiations during Experiment Phase~P\textsubscript{A}, powered by \textit{GPT-4.1}. The NSC Agents represent three stakeholders, Media-Flex, Factory-Ops, and IoT-Sense, in the \textsc{Agoran} marketplace. The agents negotiate over the three Pareto-optimal offers generated by the multi-objective optimizer (Table~\ref{tab:pareto_kpi_exp1}), ultimately reaching consensus on \emph{Offer~2} as the most balanced solution. Agreement is achieved in a \emph{single round}, excluding the initial expression of intent.}
    \label{fig:nego-snap}
\end{figure*}

Across all four phases, the three tenant agents and the Mediator Agent consistently converged to the same offer within a single JSON negotiation round, excluding the initial round request and starting from the mediator’s first response. Offer~\textit{2*} was ultimately selected in Phase~P\textsubscript{A}, with its underlying resource allocation detailed in Table~\ref{tab:pareto_res_exp1}. The complete negotiation transcript for this round is shown in Figure~\ref{fig:nego-snap}.
Table~\ref{tab:agreed_kpis} summarizes the KPIs of the agreed SLAs for each phase, as accepted by all agents; every slice constraint is satisfied in every phase (comparing Table~\ref{tab:agreed_kpis} to Table~\ref{tab:constraints}), despite diverging slice objectives and the significant MCS degradation observed in Phase~P\textsubscript{B}.

\begin{table}
\scriptsize
\centering

\begin{tabular}{p{0.2cm}p{2cm}p{0.75cm}p{0.65cm}p{0.65cm}p{0.65cm}p{0.65cm}}
\hline
\textbf{Exp} & \textbf{Application} & \textbf{Intent}  & \textbf{Tput} & \textbf{Lat.} & \textbf{Cost} & \textbf{Energy} \\
\hline
\multirow{3}{*}{PA} & Media-Flex  & eMBB  & \textbf{60} & 5.7 & €63 & 10.8\,W \\
& Factory-Ops & URLLC & 35 & \textbf{1.5} & €132 & 12.1\,W \\
& IoT-Sense & mMTC  & 38 & 5.5 & \textbf{€2} & 0.0\,W \\
\hline
\multirow{3}{*}{PB} & Media-Flex & eMBB  & \textbf{11} & 8.8 & €84 & 3.8\,W \\
& Factory-Ops & URLLC & 7  & \textbf{3.4} & €27 & 2.5\,W \\
& IoT-Sense & mMTC  & 7 & 7.5 & \textbf{€0} & 2.5\,W \\
\hline
\multirow{3}{*}{PC} & Media-Flex & eMBB  & 0 & $\infty$ & €0 & 0\,W \\
& Factory-Ops & URLLC & 0 & $\infty$ & €0 & 0\,W \\
& IoT-Sense & mMTC  & 0 & $\infty$ & €0 & 0\,W \\
\hline
\multirow{3}{*}{PD} & Media-Flex & \textbf{URLLC} & 38 & \textbf{1.5} & €2.7 & 1.7\,W \\ 
& Factory-Ops & \textbf{eMBB}  & \textbf{58} & 5.7 & €3 & 26.0\,W \\
& IoT-Sense & mMTC  & 38 & 5.5 & \textbf{€0} & 1.3\,W \\
\hline
\end{tabular}
\caption{KPIs of the \emph{agreed} SLA for each experiment phase. These SLAs reflect the capabilities of our testbed, which employs an OAI gNB with $40$\,MHz bandwidth and supports a maximum overall throughput of $133.7$\,Mbps.}
\label{tab:agreed_kpis}
\end{table}

\begin{figure}
  \centering
  \subfloat[Throughput of the three slices]{%
        \includegraphics[width=0.7\linewidth]{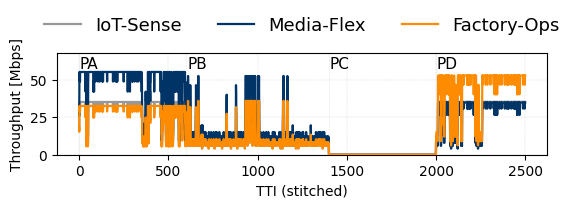}}\par
  \subfloat[Latency of the three slices]{%
        \includegraphics[width=0.7\linewidth]{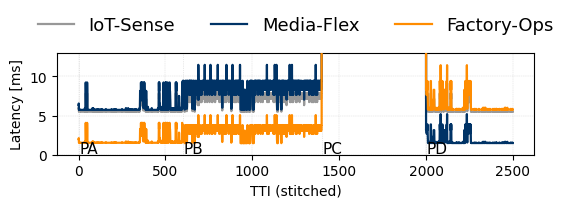}}
  \caption{Negotiated 5G slice performance across the four
           experimental phases P\textsubscript{A}\,--\,P\textsubscript{D}.}
  \label{fig:timeline}
\end{figure}

\subsection{Throughput, Latency, and Resource Use}
Figure~\ref{fig:timeline} shows the per-slice throughput and latency over 2,500 stitched Transmission Time Intervals (TTIs).

\textit{Adaptivity.} In Phase~P\textsubscript{A}, the eMBB slice (Media-Flex) achieves over $60$\,Mbps, while the URLLC latency of the Factory-Ops slice remains below $2$\,ms. When spectral efficiency collapses (Phase~P\textsubscript{B}), the agents renegotiate reduced but still feasible performance targets.
In Phase~P\textsubscript{C}, the stakeholders jointly request a power-off window, resulting in zero throughput and negligible energy consumption. Finally, in Phase~P\textsubscript{D}, the intent swap is honored: Factory-Ops is prioritized for eMBB, reaching approximately $60$\,Mbps, while Media-Flex latency is held around $1.5$\,ms.

\begin{figure}
  \centering
  \subfloat[Factory-Ops throughput]{%
        \includegraphics[width=0.6\linewidth]{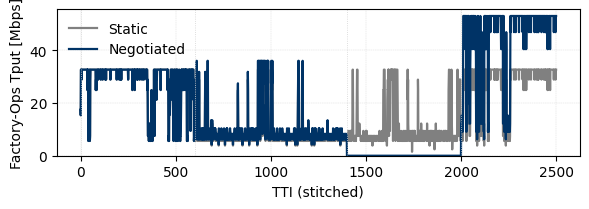}}\par
  \subfloat[Factory-Ops PRB utilization]{%
        \includegraphics[width=0.6\linewidth]{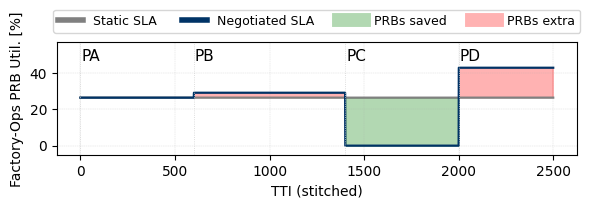}}\par
  \subfloat[Media-Flex latency]{%
        \includegraphics[width=0.6\linewidth]{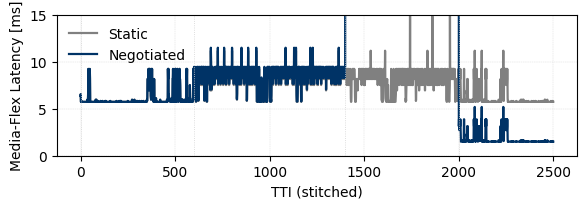}}
  \caption{Negotiated versus static SLA. Dynamic reallocation increases throughput, releases unused PRBs (green), and allocates additional PRBs only when beneficial (red), while achieving lower latency when Media-Flex becomes URLLC-critical in Phase~P\textsubscript{D}.}
  \label{fig:comparison}
\end{figure}

Figure~\ref{fig:comparison}b shows physical PRB utilization for Factory-Ops, comparing \textsc{Agoran}'s dynamic negotiation with conventional static configurations. Dynamic allocation reduces PRB usage by 15,888~PRBs (24\%) during periods of low demand and adds only 10,422~PRBs (16\%) during high-demand intervals. This results in an overall net PRB saving of 8.3\% over the full trace.
Moreover, this flexible allocation enables targeted, on-demand improvements in Factory-Ops throughput, as shown in Figure~\ref{fig:comparison}a. When the slice switches to eMBB priority in Phase~P\textsubscript{D}, throughput increases by up to $66$\%—from 35\,Mbps in Phase~P\textsubscript{A} to $58$\,Mbps in Phase~P\textsubscript{D}.
Finally, Figure~\ref{fig:comparison}c compares the latency of the Media-Flex slice under negotiated and static SLAs. Agentic negotiation reduces the median RTT by $73.4$\%, from $5.7$\,ms to $1.5$\,ms, once the slice transitions to URLLC.

\subsection{Discussion and Takeaways}
\begin{itemize}
\item \textit{One-Round Consensus.} The tripartite agent design consistently reaches unanimous agreement after a single message exchange, despite conflicting stakeholder priorities and dynamic radio conditions.
\item \textit{Constraint Satisfaction \& Flexible QoS Gains.} Negotiated SLAs always satisfy slice-level constraints while enabling flexible QoS improvements: aggregate throughput increases by up to 66\%, and URLLC latency is reduced by up to 73.4\%, compared to the static baseline.
\item \textit{Spectrum Efficiency.} Closed-loop control achieves a net 8.3\% PRB saving over the full trace, demonstrating more efficient and targeted resource allocation.
\item \textit{Human-Friendly Ultra-Flexibility.} Stakeholders articulate rich, natural-language intents (e.g., “AR e-sports all night”) rather than relying on rigid slice templates. The optimizer and agents translate these into quantitatively optimal, regulation-compliant resource directives. The negotiation framework remains agile with respect to both evolving stakeholder intents and time-varying channel conditions.
\end{itemize}

In summary, this use case validates \textsc{Agoran} as a fully autonomous yet human-centric marketplace that reconciles high-level business intent with radio resource constraints, adapts through non-real-time control loops, and enhances both QoS and spectrum efficiency, an essential capability for ultra-flexible 6G deployments.

\section{Limitations and Future Work}
\label{limitations}
This study presents the first full-stack demonstration of an agentic marketplace for RAN automation. However, several limitations remain. 

First, the evaluation is conducted on a \emph{single-site} OAI/FlexRIC deployment with three slices and three UEs.  
While the channel traces and intent dynamics are realistic, this setup cannot surface scalability challenges that may emerge in multi-gNB, large-scale service environments involving thousands of agents. 

Second, the current optimizer considers four KPIs (throughput, latency, cost, energy) and four resource types.  
In contrast, real-world networks must manage dozens of performance metrics and navigate cross-domain constraints spanning transport, edge, and core infrastructure. 

Third, hardware and API limitations, specifically, one on-premises A6000 GPU and GPT-4.1 usage quotas, restricted us to models with 70B parameters or fewer and a fine-tuning dataset of 100 synthetic dialogues. Larger open-weight checkpoints or deeper pre-training could help close the remaining performance gap. 

Finally, the retrieval pipeline and Trust Score logic are intentionally lightweight by design. Future work could incorporate advanced techniques such as negative sampling, cross-encoders, or probabilistic calibration to improve observability precision and detect adversarial behavior more reliably.

These limitations also define our next steps. Ongoing work aims to scale the optimizer and agent framework to multi-gNB clusters,  
integrate transport and core network objectives, and explore the distillation of 8–13\,B models into sub-second micro-agents.  
On the governance side, we are extending the Trust Score with probabilistic guarantees and differential privacy hooks. We release a subset of the artifacts, including code, traces, and fine-tuning notebooks, to encourage the community to stress-test and extend \textsc{Agoran} toward truly large-scale, stakeholder-centric 6G networks.

\section{Conclusion}
\label{sec:conclusion}

We introduced \textsc{Agoran}, the first tripartite agentic-AI framework that empowers business stakeholders to express network intents in natural language, enabling dedicated LLM agents to negotiate, arbitrate, and enforce them in real time. Leveraging a live Open and AI RAN testbed, we demonstrated that: (i) a watcher-driven vector store effectively grounds agents in real-time telemetry without requiring continuous polling; (ii) a rule-based Trust Score gate mitigates hallucinations and manipulation; (iii) LLM-based incentive mechanisms can restore trust in malicious agents; and (iv) a micro-fine-tuned $1$B-parameter LLaMA model achieves approximately $80\,\%$ of GPT-4.1’s decision quality in multi-agent SLA negotiations, while requiring only $6\,\mathrm{GiB}$ of memory and converging within $1.3\,\mathrm{s}$.

End-to-end RAN-slice experiments validate the practical benefits of intent-driven negotiation: aggregate throughput increases by $37\,\%$, URLLC latency decreases by $73\,\%$, and spectrum utilization drops by $8.3\,\%$ relative to a static baseline. Moreover, agile one-round consensus is preserved even under channel fading and evolving business priorities. These results position \textsc{Agoran} as a concrete, standards-compliant step toward ultra-flexible, stakeholder-centric 6G networks, and open new research directions in agentic observability, lightweight agent distillation for network-specific tasks such as SLA negotiation, and cross-domain intent reconciliation. An illustrative live demo is shown \href{https://www.youtube.com/watch?v=h7vEyMu2f5w&ab_channel=BubbleRAN}{$https://www.youtube.com/watch?v=h7vEyMu2f5w\&ab_channel=BubbleRAN$}

\section*{Ethical and Safety Considerations}
\label{ethical}

\textbf{Regulatory alignment.}  
\textsc{Agoran} is explicitly designed to operate under binding policy constraints. Its Legislative branch grounds every decision in formal regulatory documents (e.g., 3GPP, ETSI, and national spectrum rules), ensuring \emph{de jure} compliance while facilitating alignment and conformity with emerging frameworks such as the EU AI Act \cite{eu_ai_act} and NIST’s AI Risk Management Framework \cite{nist_ai_rmf}.

\textbf{Transparency and auditability.}  
All agent messages should include a cryptographic signature (e.g., HMAC \cite{hmac_wiki}), a natural-language rationale, and a numeric Trust Score derived from verifiable rules. Together, these elements form an immutable audit trail that regulators or third-party auditors can replay to fully reproduce every allocation and sanction.

\textbf{Bias and hallucination control.}  
The rule-based Trust Score filters toxic or deceptive content and
down-weights out-of-distribution hallucinations before they can influence resource policy. As a non-parametric and fully
interpretable mechanism, the filter can be inspected and adjusted by human experts, an essential requirement for AI systems operating in network-critical environments.

\textbf{Privacy and data minimization.}  
Watcher streams export only coarse-grained counters and configuration
deltas, never user payload. All embeddings stored in the vector database are ephemeral and perturbed with sub-token noise, mitigating the risk of membership inference attacks while still enabling effective semantic search.

\textbf{Alignment incentives.}  
The Judicial branch combines real-time scoring with proportional
rewards and penalties, guiding agents toward cooperative equilibria.
Future work will explore the integration of differential privacy budgets and cryptographic proofs-of-honesty to reinforce these incentives at scale.

\textbf{Responsible roadmap.}  
We view \textsc{Agoran} as a \emph{safety-critical stepping stone}:
while agentic negotiation enhances flexibility, unchecked autonomy could introduce systemic risks.  Accordingly, we release code under a
responsible licensing model, disclose all training data sources, and
invite cross-domain red-teaming prior to any production deployment.

By embedding transparency, policy grounding, and behavioral safeguards into its core architecture, \textsc{Agoran} aims to realize the benefits of agentic AI for 6G networks, while respecting the ethical, legal, and societal boundaries that such powerful technology demands.

\section*{CRediT authorship contribution statement}
\textbf{Ilias Chatzistefanidis:} Conceptualization, Methodology, Software, Validation, Formal analysis, Investigation, Resources, Data Curation, Writing - Original Draft, Writing - Review \& Editing, Visualization, Supervision, Project administration. 

\textbf{Navid Nikaein:} Conceptualization, Methodology, Resources, Writing - Original Draft, Writing - Review \& Editing, Visualization, Supervision, Project administration, Funding acquisition. 

\textbf{Andrea Leone:} Conceptualization, Methodology, Software, Validation, Formal analysis, Investigation, Resources, Data Curation, Writing - Original Draft, Visualization. 

\textbf{Ali Maatouk:} Methodology, Validation, Formal analysis, Investigation, Writing - Original Draft, Visualization.

\textbf{Leandros Tassiulas:} Methodology, Validation, Formal analysis, Investigation, Writing - Original Draft, Supervision.

\textbf{Roberto Morabito:} Methodology, Investigation, Writing - Original Draft, Supervision.

\textbf{Ioannis Pitsiorlas:} Methodology, Software, Validation, Formal analysis, Investigation, Writing - Original Draft, Visualization.

\textbf{Marios Kountouris:} Methodology, Validation, Formal analysis, Investigation, Writing - Original Draft, Visualization, Supervision.

\section*{Declaration of competing interest}

The authors declare that they have no known competing financial interests or personal relationships that could have influenced the work reported in this paper.

\section*{Data Availability}

To aid reproducibility and foster further exploration, we will provide a concise set of artifacts:

\begin{itemize}
    \item \textbf{Demo video.} A short proof-of-concept showing three stakeholder slices negotiating over the live OAI/FlexRIC testbed. Watch online \href{https://www.youtube.com/watch?v=h7vEyMu2f5w&ab_channel=BubbleRAN}{\textcolor{blue}{here}}.
    \item \textbf{Open-source repository (upon acceptance).} A repository containing RAN traces, part of the agent code, the trust framework, training data, and optimization techniques. This resource is intended to support further research and innovation in the field.
\end{itemize}

\section{Acknowledgment}
This work was supported by the European Commission as part of the Horizon Europe 2022 6Green, ADROIT6G, 6G-Cloud, 6G-INTENSE, ROBUST-6G, and 6G-LEADER Projects under Grant Agreements No.101096925, 101095363, 101139073, 101139266, 101139068, and 101192080, respectively.

\bibliographystyle{elsarticle-num}   
\bibliography{biblio}

\end{document}